\input epsf.sty
\overfullrule=0pt
\font\hbf=cmbx10 scaled\magstep2
\font\bbf=cmbx10 scaled\magstep1

\font\caps=cmcsc10
\font\sbf=cmbx8
\font\srm=cmr8
\font\eightpoint=cmr8
\font\fivepoint=cmr5
\font\eightit=cmti8
\font\eightbf=cmbx8

\def\A{{\cal A}}
\def\G{{\cal G}}
\def\CB{{\cal B}}

\def\H{{\cal H}}

\def\M{{\cal M}}
\def\N{{\cal N}}
\def\O{{\cal O}}
\def\K{{\cal K}}
\def\W{{\cal W}}
\def\R{{\bf R}}
\def\BC{{\bf C}}
\def\I{{\rm i}}
\def\E{{\rm e}}
\def\sabsatz{\par\smallskip\noindent}
\def\mabsatz{\par\medskip\noindent}
\def\babsatz{\par\vskip 1cm\noindent}

\def\newline{\hfil\break\noindent}

\def\Box{{\vcenter{\vbox{\hrule width4.4pt height.4pt 
              \hbox{\vrule width.4pt height4pt\kern4pt\vrule width.4pt}
	       \hrule width4.4pt }}}}
\def\norm#1{\|#1\|}
\def\frac#1#2{{#1\over#2}}
\def\B1{{\mathchoice {\rm 1\mskip-4mu l} {\rm 1\mskip-4mu l}
  {\rm 1\mskip-4.5mu l} {\rm 1\mskip-5mu l}}}
\def\BC{{\mathchoice {\setbox0=\hbox{$\displaystyle\rm C$}\hbox{\hbox
  to0pt{\kern0.4\wd0\vrule height0.9\ht0\hss}\box0}}
  {\setbox0=\hbox{$\textstyle\rm C$}\hbox{\hbox
  to0pt{\kern0.4\wd0\vrule height0.9\ht0\hss}\box0}}
  {\setbox0=\hbox{$\scriptstyle\rm C$}\hbox{\hbox
  to0pt{\kern0.4\wd0\vrule height0.9\ht0\hss}\box0}}
  {\setbox0=\hbox{$\scriptscriptstyle\rm C$}\hbox{\hbox
  to0pt{\kern0.4\wd0\vrule height0.9\ht0\hss}\box0}}}}
  \headline={\hfill{\fivepoint HJBJY--May 7 1998}}
\magnification=\magstep1
\centerline{\hbf Modular Groups of Quantum Fields} 
\mabsatz
\centerline {\hbf in Thermal States}
\vskip 0.4cm
\centerline{\caps H.J. Borchers}
\sabsatz
\centerline{Institut f\"ur Theoretische Physik} 
\centerline{Universit\"at G\"ottingen}
\centerline{Bunsenstrasse 9, D 37073 G\"ottingen}
\centerline{and}
\centerline{\caps J. Yngvason}
\sabsatz
\centerline{Institut f\"ur Theoretische Physik} 
\centerline{Universit\"at Wien}
\centerline{Boltzmanngasse 5, A 1090 Wien}

{\narrower \sabsatz
{\sbf Abstract:}\newline{\srm
For a quantum field in a thermal equilibrium state we discuss the 
group generated by time translations and the modular action associated 
with an algebra invariant under half-sided translations.  The modular 
flows associated with the algebras of the forward light cone and a 
space-like wedge admit a simple geometric description in two 
dimensional models that factorize in light-cone coordinates.  At large 
distances from the domain boundary compared to the inverse temperature 
the flow pattern is essentially the same as time translations, whereas 
the zero temperature results are approximately reproduced close to the 
edge of the wedge and the apex of the cone.  Associated with each 
domain there is also a one parameter group with a positive generator, 
for which the thermal state is a ground state.  Formally, this may be 
regarded as a certain converse of the Unruh-effect.

}\sabsatz}
\mabsatz

\noindent
{\bbf 1. Introduction}
\mabsatz

Algebraic quantum field theory in the sense of Araki, Haag, Kastler 
[Ha96] is concerned with von Neumann 
algebras $\M(\O)$ of observables localized in space time domains $\O$, 
together with states $\omega$ on these algebras satisfying some 
physical selection criterion.  Due to the 
Reeh-Schlieder property of quantum field theory [RS61] one may associate 
with certain regions $\O$ and states $\omega$ the Tomita-Takesaki 
modular objects $\Delta_{\O,\omega}$ and $J_{\O,\omega}$ [Ta70], 
[KR86].  The 
positive operator $\Delta_{\O,\omega}$ generates a one parameter group 
$\hbox{\rm ad}\,\Delta_{\O,\omega}^{\I t}$ of 
automorphisms of $\M(\O)$, and the conjugation $\hbox{\rm ad}\, 
J_{\O,\omega}$ defined by the antiunitary $J_{\O,\omega}$ maps 
$\M(\O)$ onto its commutant in the GNS Hilbert space corresponding to 
$\omega$.

Important structural properties of the theory are encoded in the modular 
objects, see, e.g., [BDL90], [Bch95], but an explicit description of 
$\Delta_{\O,\omega}^{\I t}$ and $J_{\O,\omega}$
has so far only been obtained in the following 
cases with $\omega$ a vacuum state:

\item {(a)}$\O$ is a space like wedge and the local algebras are 
generated by Wightman fields that transform covariantly with a 
finite dimensional representation of the Lorentz group [BW75,76].

\item {(b)}$\O$ is a  forward light cone and $\M(\O)$ is generated by  a 
massless, non-interacting field [Bu78]. 

\item {(c)}$\O$ is a double cone and $\M(\O)$ is generated by  
conformally
covariant fields [HL82]. 

\item {(d)} $\O$  is a space like wedge and the local algebras are 
generated 
by generalized free fields of a certain type which break Lorentz 
covariance [Y93].

In case (a) the modular group is the group of Lorentz boosts that 
leave the wedge invariant, and the conjugation is the PCT operator 
(combined with a rotation).  Cases (b) and (c) can be reduced to case 
(a) by a conformal mapping onto the wedge of the forward light cone 
and the double cone respectively.  In (b) the modular group is the 
dilation group, and in (c) it consists of the conformal 
transformations that leave the double cone invariant.  In the examples 
considered in (d) the action of the modular group is in general 
non-local, i.e., an algebra $\M(\O_{1})$ with $\O_{1}$ a bounded 
subset of the wedge need not be mapped into an $\M(\O_{2})$ with 
$\O_{2}$ bounded.

A key to a general understanding of possible geometric 
interpretations of modular groups is the interplay between the modular 
action and certain subgroups of the space-time translations.  In 
[Bch92] it was shown that the modular group of a space-like wedge in a 
vacuum state acts on the translation group like the Lorentz boosts 
that leave the wedge invariant.  Subsequently Wiesbrock [Wie93] 
introduced the concept of a half-sided modular inclusion and proved a 
certain converse of the results of [Bch92], namely that the 
two-dimensional translation group can be recovered from the modular 
groups of the wedge and some of its translates.

In this paper we want to investigate the modular groups when $\omega$ 
is a thermodynamic equilibrium state (KMS state) rather than a vacuum 
state.  In the next Section 2 we discuss the generalizations of the 
results of [Bch92] to KMS states.  We investigate the commutation 
relations between the time translations and the modular group in a KMS 
state for any domain that is mapped into itself under half-sided time 
translations.  Using the results of [Bch95] we prove that the time 
translations and modular action together give rise to a representation 
of the abstract Lie group generated by one dimensional dilations and 
translations.  The important observation that half-sided modular 
actions always lead to a representation of this group was first made 
by Wiesbrock [Wie93], [Wie97].  We express all its one parameter 
subgroups in terms of the translations and the modular group.  Of 
particular interest is a subgroup with a positive generator.  This 
group acts on the global observable algebra for positive values of the 
group parameter.

The group relations alone do not determine the modular action and the 
group with positive generator, but they put definite restrictions on 
the possible disclocalization of observables by the group actions.  
More precisely, if $\N$ denotes the observable algebra of a domain 
invariant under half sided translations and $\N(t)$ its time translate 
by $t$, then the modular group $\Delta_\N^{\I u}$ of $\N$ transforms 
$\N(t)$ into $\N(\varphi(u,t))$ with a certain function 
$\varphi(u,t)$.  Likewise, the group with a positive generator 
transforms $\N(t)$ into $\N(\psi(\tau,t))$, where $\psi$ is another 
function of $t$ and the group parameter $\tau$.  The precise 
statements are given in Theorem 2.1.  We also discuss the action of 
$\Delta_\N^{\I u}$ on individual observables in $\N(t)$ for $t$ large 
and show that, in a sense made precise in Theorems 2.2 and 2.3, this 
action approximates a time translation by $-\beta u$ as $t/\beta\to\infty$.

In Section 3 we consider two dimensional models that factorize in the 
light cone coordinates.  Applying the results of the previous section 
to the algebras on each of the light rays one obtains a geometric 
description of the actions of the groups associated with the forward 
light cone and a space like wedge.  In the case of the forward light 
cone, the algebra of a translated light cone is mapped into another 
such algebra.  An analogous statement holds for the wedge.  The flow 
patterns are illustrated in Figs.\ 1-2.  Close to the apex of the 
light cone and the edge of the wedge the actions of the modular flow 
is essentially the same as for the zero temperature case, i.e., 
dilations for the forward light cone and Lorentz boosts for the wedge.  
On the other hand, at large distances from the domain boundary 
compared to the inverse temperature the modular flow approaches the 
dynamical flow, i.e., the time translations.

The one parameter unitary group with positive generator 
associated with the forward light cone, 
which in the limiting case of zero temperature reduces to time 
translations, approximates the dynamical flow close to the apex of the 
light cone. It corresponds everywhere to a decelerated movement 
towards the origin in the space variable (Fig.\ 3).  Formally at 
least, this may be regarded as a reverse Unruh-Effect [U76], [Sew80], [Sew82]: 
In the latter the vacuum appears as a KMS state with respect to a 
dynamics that accelerates points towards light-like infinity, here a 
KMS state appears as a vacuum with respect to a dynamics that moves 
points from light-like infinity towards the origin of space.

For the wedge there is also a unitary group with positive generator 
which has the KMS state as a ground state.  This group operates on the 
observables for a restricted parameter range. It approximates the time 
translations close to the space axis and light 
like translations far away from the space axis.  The action of this 
group is illustrated in Fig.\ 4.  This action may also be interpreted 
as a kind of reverse Unruh effect, because the acceleration is here 
away from the wedge, whereas in the usual Unruh effect the acceleration 
points in the direction of the wedge.

In Section 4 we compute explicitly the modular groups and the groups 
with positive generator for a quasi free KMS state on the Weyl algebra 
of a generalized free field in 2D space time that factorizes in light 
cone coordinates.  For a field of minimal scaling dimension one 
obtains a strengthening of the general results of the previous section 
on the group actions: A local algebra $\M(\O)$ with $\O$ a double cone 
is transformed into an algebra of the same kind.  For fields of higher 
scaling dimension, however, double cone localization may be 
get lost under the group action and only a localization in a 
translated light cone or wedge remains.
\babsatz
\noindent
{\bbf 2. The group generated by  translations and the modular action}
\mabsatz
Let $(\A,\alpha_{t})$ be a $C^*$-dynamical system and $\CB$ a 
subalgebra, such that
$$\alpha_{t}\CB\subset \CB\qquad\hbox{\rm for\ }t\geq 0.\eqno(2.1)$$
Suppose furthermore that the algebra
$\cup_{t\in\R}\alpha_{t}\CB$
is norm dense in $\A$. Let $\omega$ be a KMS state [BR79] for the 
dynamical system $(\A,\alpha_{t})$ at inverse temperature $\beta$ and 
denote by $\pi$ the corresponding GNS representation of  $\A$ with 
cyclic vector $\Omega$, and by 
$T(t)$ the unitary implementation of $\alpha_{t}$ on the GNS Hilbert 
space $\H$. Put $\M=\pi(\A)^{\prime\prime}$ and 
$\N=\pi(\CB)^{\prime\prime}$.

Because of the analyticity properties of the time translations in a 
KMS state the vector $\Omega$ is separating for $\M$ and hence also 
for $\N$.  Moreover, $\Omega$ is cyclic for $\M$ (by definition) and 
since $\cup_{t\in\R}\alpha_{t}\CB$ is dense in $\A$ it follows by a 
Reeh-Schlieder type argument that $\Omega$ is cyclic for $\N$ also.

Let $\Delta_{\M}$ and $J_{\M}$ be the modular objects corresponding 
to $\Omega$ and  $\M$. We have
$$\Delta_{\M}^{\I s}=T(-\beta s)\eqno(2.2)$$
where the sign is a consequence of different conventions in 
physics and mathematics: For $A\in\M$ the expression $T(t)A\Omega$
has an analytic continuation into the strip $S(0,\beta/2)$, where
$$S(a,b):=\{z\in\BC :a<\hbox{\rm Im\ }z<b\},\eqno(2.3)$$
while $\Delta_{\M}^{\I s}A\Omega$ has an analytic continuation into 
$S(-1/2,0)$, by the sign convention in modular theory.
Since $J_{\M}\Delta_{\M}^{1/2}A\Omega=A^*\Omega$, it follows from
(2.2) that
$$T(t+i\beta)A\Omega=J_{\M}T(t)A^*\Omega.\eqno(2.4)$$
By assumption (2.1) we have 
$$T(t)\N T(-t)\subset\N\qquad\hbox{\rm for }t\geq 0,\eqno(2.5)$$
i.e., we are in the situation of a half-sided translation in the sense 
of [Bch92]. Because of (2.2) we are also in the situation of a half-sided 
modular inclusion in the sense of Wiesbrock [Wie93], i.e.,
$$\Delta_{\M}^{\I s}\N\Delta_{\M}^{-\I  s}\subset\N\qquad\hbox{\rm for }s
\leq 0.\eqno(2.6)$$
If $T(t)$ had a positive generator, then (2.5) would imply the well 
known relations [Bch92] between $T(t)$ and the modular group 
$\Delta_{\N}^{\I u}$.  In a KMS state, however, the spectrum of 
the Hamiltonian is the whole real axis and the analysis of  [Bch92]
has to be generalized. The main results of this generalization are stated 
in
Eqs.\ (2.20), (2.29) and (2.31) below.

We start with a heuristic discussion of the consequences of (2.6), 
similar to that in [Wie93].  This discussion disregards questions of 
domains of unbounded operators, but it leads quickly to the 
commutation relations between $\Delta_{\M}^{\I s}$ and 
$\Delta_{\N}^{\I u}$ stated in [Wie97].  A rigorous proof of these 
relations follows from the results of [Bch95] and will be given after 
the discussion.

Since $\N\subset \M$ it follows by standard arguments that 
$\Delta_{\N}\geq \Delta_{\M}$ and this, domain questions aside,
implies that 
$$G:=\log\,\Delta_{\N}-\log\,\Delta_{\M}\eqno(2.7)$$
is a non-negative operator, because log is an operator monotone 
function. Eqs.\ (2.1), (2.2), (2.6) and the Trotter product formula now lead to
$$e^{\I \tau G}\N e^{-\I  \tau G}\subset\N\qquad\hbox{\rm for }
\tau\geq 0.\eqno(2.8)$$
Putting $U(\tau):=\exp(\I\tau G)$, Eq.\ (2.8) and $G\geq 0$ imply
[Bch92]
$$\Delta_{\N}^{\I u}U(\tau)\Delta_{\N}^{-\I  u}=U(e^{-2\pi 
u}\tau)\eqno(2.9)$$
for all $\tau,u\in\R$. Hence we obtain a unitary representation of the 
two parameter Lie group $\G$ with elements $(\tau,u)\in\R^2$ and the 
composition law
$$(\tau,u)\circ (\tau',u')=(\tau+e^{-2\pi u}\tau', u+u').\eqno(2.10)$$
The representation $U(\tau,u)$ corresponding to (2.9) is 
$$U(\tau,u):=e^{\I \tau G}\Delta_{\N}^{\I u}.\eqno(2.11)$$

The group $\G$ defined by (2.10) is the semidirect product of $\R$ with 
itself and is the unique two dimensional non-abelian Lie group 
(\lq\lq$ax+b$-group"). 
Some of its properties are discussed in [Bch98]. 

For a discussion of the one parameter subgroups and the Lie algebra 
of $\G$ it is convenient to realize the group in terms of $2\times 2$ 
matrices:
$$(\tau,u)\leftrightarrow\pmatrix{
1&\tau\cr 0&1\cr}\cdot \pmatrix{
e^{-2\pi u}&0\cr 0&1\cr}=\pmatrix{
e^{-2\pi u}&\tau\cr 0&1\cr}.\eqno(2.12)$$
It is straightforward to determine the one parameter subgroups, 
$r\mapsto g(r)$ of $\G$. These have the form
$$g_{a,b}(r)=\pmatrix{
e^{ar}&{b\over a}(e^{ar}-1)\cr 0&1\cr}\eqno(2.13)$$
with $a,b\in\R$. In the half-plane $(\tau,e^{-2\pi u})\in\R\times\R_{+}$
these correspond to straight lines through $(0,1)$. The 
infinitesimal generator of $g_{a,b}(r)$ is 
$$\hat g_{a,b}=\left .{d\over dr}g_{a,b}(r)\right|_{r=0}=\pmatrix{
a&b\cr 0&0\cr}.\eqno(2.14)$$
The group $\Delta_{\N}^{\I u}$ corresponds to $a=-2\pi$, $b=0$; the 
group $\exp(i\tau G)$ to $a=0$, $b=1$. Since the generator of 
$\Delta_{\M}^{\I s}$ is $\log \Delta_{\N}-G$, this one parameter group 
corresponds to $a=-2\pi$, $b=-1$. Denoting for short the one 
parameter subgroups of $\G$ in these three cases by $g_{\N}(u)$, $g_{\rm 
pos}(\tau)$ 
and 
$g_{\M}(s)$ respectively, we have
$$g_{\N}(u)=\pmatrix{
e^{-2\pi u}&0\cr 0&1\cr},\quad g_{\rm pos}(\tau)=\pmatrix{
1&\tau\cr 0&1\cr},\quad g_{\M}(s)=\pmatrix{
e^{-2\pi s}&{1\over 2\pi}(e^{-2\pi s}-1)\cr 0&1\cr}.\eqno(2.15)$$
One verifies the relation
$$g_{\N}(u)\cdot g_{\M}(s)=g_{\M}(F(u,s))\cdot g_{\N}(-F(u,s)+s+u)
\eqno(2.16)$$
with
$$F(u,s)=-{1\over 2\pi}\log\left\{1+e^{-2\pi u}(e^{-2\pi
s}-1)\right\},\eqno(2.17)$$
provided
$$1+e^{-2\pi u}(e^{-2\pi 
s}-1)>0\eqno(2.18)$$
which is always fulfilled for $s\leq 0$.  The relation corresponding 
to (2.16) for the  modular groups $\Delta_{\N}^{\I u}$ and 
$\Delta_{\M}^{\I s}$ 
is
$$\Delta_{\N}^{\I u}\cdot\Delta_{\M}^{\I s}=\Delta_{\M}^{\I F(u,s)}
\cdot \Delta_{\N}^{\I (-F(u,s)+s+u)}.\eqno(2.19)$$
This relation appears also in [Wie97]; our heuristic discussion has
brought its group theoretical origin into focus.

In terms of the original translation group $T(t)$ we can, because 
of (2.2), write (2.19) as
$$\Delta_{\N}^{\I u}\cdot T(t)\cdot\Delta_{\N}^{-\I  u}
=T\left({\beta\over 2\pi}
\log\left\{1+e^{-2\pi u}(e^{2\pi t/\beta }-1)\right\}\right)
\cdot \Delta_{\N}^{\I {1\over 2\pi}
\log\left\{1+e^{-2\pi u}(e^{2\pi t/\beta }-1)\right\}-\I{t\over 
\beta}}.\eqno(2.20)$$
In the limit $\beta\to \infty$ we recover the Bisognano-Wichmann result
$$\Delta_{\N}^{\I u}\cdot T(t)\cdot\Delta_{\N}^{-\I  u}=T\left(e^{-2\pi 
u}t\right).\eqno(2.21)$$

\sabsatz
The 
one parameter groups $g_{a,b}(r)$ can be expressed 
in terms of $g_{\M}(s)$ and 
$g_{\N}(u)$ :
$$g_{a,b}(r)=g_{\M}(s(r))\cdot g_{\N}(u(r))=g_{\N}(u(-r))\cdot 
g_{\M}(s(-r))\eqno(2.22)$$
with
$$\eqalign{s(r)&=-{1\over 2\pi}\log \left\{1+{2\pi b\over a}
(e^{ar}-1)\right\}\cr u(r)&=-s(r)-{a\over 2\pi}r.\cr}\eqno(2.23)$$
Specializing to $a=0$, $b=1$ we obtain for $\tau>-1/(2\pi)$
$$g_{\rm pos}(\tau)=g_{\M}\left(-(2\pi)^{-1}\log(1+2\pi \tau)\right)\cdot
g_{\N}\left((2\pi)^{-1}\log(1+2\pi \tau)\right),\eqno(2.24)$$
and hence
$$U(\tau)=\Delta_{\M}^{-{\I \over 2\pi} \log(1+2\pi \tau)}  
\cdot \Delta_{\N}^{{\I \over 2\pi} \log(1+2\pi \tau)}=\Delta_{\N}^{-{\I \over 
2\pi}\log(1-2\pi \tau)} \cdot 
\Delta_{\M}^{{\I \over 2\pi} \log(1-2\pi \tau)}.\eqno(2.25)$$
The first representation can be used for $\tau>-1/(2\pi)$, the second one 
for
$\tau<1/(2\pi)$.

The group $g_\N(u)$ operates on $g_{\rm pos}(\tau)$ according to
$$g_\N(u)g_{\rm pos}(\tau)g_\N(-u)=g_{\rm pos}(\exp(-2\pi 
u)\tau),\eqno(2.26)$$
which is just the abstract form of the basic relation (2.9).
This is a special case of the general relation 
$$g_{a,b}(r)g_{\rm pos}(\tau)g_{a,b}(-r)=g_{\rm 
pos}(\exp(ar)\tau).\eqno(2.27)$$
For $a=-2\pi$, $b=-1$, i.e. $g_{\M}$, the corresponding relation for the 
unitary 
groups on 
Hilbert space is
$$\Delta_{\M}^{\I s}U(\tau)\Delta_{\M}^{-\I s}=
U(\exp(-2\pi s)\tau),\eqno(2.28)$$
which follows also directly from (2.25) and (2.19). 
We note in passing that (2.28) may 
be interpreted as an 
``Anosov relation'' that
leads  to exponential clustering of matrix elements of 
the time translations $T(t)=\Delta_\M^{-\I t/\beta}$ in states of the form 
$A\Omega$ with $A$ in a dense subalgebra of $\M$ 
[ENTS95].

Defining $\Gamma(\tau):=U(\tau/\beta)$ we have by (2.25)
$$\eqalign{\Gamma(\tau)&=T\left({\beta\over 
2\pi}\log\{1+(2\pi\tau/\beta)\}\right)
\cdot \Delta_{\N}^{{\I \over 2\pi} \log\{1+(2\pi 
\tau/\beta)\}}\cr &=\Delta_{\N}^{-{\I \over 2\pi} \log\{1-(2\pi 
\tau/\beta)\}}\cdot T\left(-{\beta\over 
2\pi}\log\{1-(2\pi\tau/\beta)\}\right),\cr}\eqno(2.29)$$
where the first equality is valid for $\tau>-\beta/(2\pi)$ and the second for 
$\tau<\beta/(2\pi)$.
Evidently $\Gamma(\tau)\to T(\tau)$ for $\beta\to\infty$, and 
$$G/\beta=H+{1\over\beta}\log\, \Delta_{\N}\eqno(2.30)$$
tends in this limit to the Hamiltonian $H$, which in the vacuum 
representation 
is 
$\geq 0$.

The relation (2.28) means that
$$T(t)\Gamma(\tau)T(-t)=\Gamma(\exp(2\pi t/\beta)\tau).\eqno(2.31)$$
By (2.8) and our  assumption that $\cup_t \alpha_t{\cal B}$ is norm dense 
in ${\cal A}$ (and hence $\cup_t {\rm ad}\, T(t)\N$ weakly dense in ${\cal 
M}$), 
we may thus conclude that
$${\rm ad}\,\Gamma(\tau)\M\subset \M \qquad\hbox{ for all $\tau\geq 
0$}.\eqno(2.32)$$

\sabsatz

A rigorous proof of the relations (2.19) and (2.25) (and hence of 
(2.20), (2.29)
and (2.31)) can be obtained by applying Theorems A and B in [Bch95] to 
the operator valued functions
$$V(v)=\Delta_{\M}^{-\I  v}\Delta_{\N}^{\I v}\eqno(2.33)$$
and $W(w)=V(v(w))$, where
$$v(w)={1\over 2\pi}\log\left(1+e^{2\pi w}\right).\eqno(2.34)$$
The function $V(v)$ has a bounded analytic continuation into the 
strip $S(0,1/2)$ with continuous boundary values and satisfies the 
relation
$$V\left(v+{\I \over 2}\right)=J_{\M}V(v)J_{\N}\eqno(2.35)$$
for $v\in\R$. Moreover, ad $V(v)$ maps $\N$ into $\N$ for $v\geq 0$ 
and the commutant $\N'$ into $\N'$ for $v\leq 0$. By (2.35) it follows 
that
ad $V\left(v+{\I \over 2}\right)$ maps $\N'$ into $\N'$ for all $v$.

In order to apply Theorem B in [Bch95] we have to map the strip $S(0,1/2)$ 
biholomorphically onto itself in such a way that $\R$ is mapped onto 
$\R_{+}$ and $\R+{\I \over 2}$ onto $\left(\R+{\I \over 
2}\right)\cup\R_{-}$.
The map (2.34) accomplishes this. It has a singularity at $w=\I/2$, but 
as 
remarked in [Bch95] such a singularity is harmless.

[The reason is as follows: Theorem B in [Bch95] is based on 
the edge-of-the-wedge theorem, applied to matrix elements of the operator 
valued function 
$$(u,w)\mapsto \Delta_{\N}^{\I u}W(w)\Delta_{\N}^{-\I  u}.\eqno(2.36)$$
These matrix elements have bounded analytic continuations, which are 
continuous at the boundary of their domain with the possible 
exception of points with $w=\I/2$. By the dominated convergence 
theorem and the boundedness of (2.36) this piece-wise continuity is 
sufficient 
to ensure coincidence 
of boundary values in the sense of distributions. The 
edge-of-the-wedge theorem then implies analyticity in the coincidence 
region, so continuity in the points with $w=\I/2$ holds a fortiori.]

Theorem B in [Bch95] leads to the general relations
$$\Delta_{\N}^{\I u}W(w)\Delta_{\N}^{-\I  u}=W(w-u)\eqno(2.37)$$
and
$$J_{\N}W(w)J_{\N}=W\left(w+{\I\over 2}\right).\eqno(2.38)$$
Eq.\ (2.37) is precisely (2.19) in case (2.18) holds, but note that 
(2.37) is 
true for 
all $u,w\in\R$. As noted by Wiesbrock ([Wie93], [Wie97]) these relations 
imply that $\Delta_{\N}^{\I u}$ and $\Delta_{\M}^{\I s}$
generate a unitary representation of the Lie group $\G$. The 
infinitesimal generators $\log\Delta_{\N}$ and $\log\Delta_{\M}$, 
together with their real linear combinations, are thus essentially 
self adjoint on a common core.
The representation $U(\tau)$ of the one parameter subgroup
$g_{\rm pos}(\tau)$ fulfills together with $\Delta_{\N}^{\I u}$ the relation 
(2.9)
(because of the corresponding relation in $\G$), and this implies by 
[Wie92] that $U(\tau)=\exp(\I\tau G)$ with $G\geq 0$. Hence the starting 
point of the heuristic discussion is rigorously justified.

The following theorem summarizes the main conclusions of the preceding 
discussion and states in addition the most important consequence of 
the relations (2.20), (2.29) and (2.31) for the present investigation, 
namely the action of the group $\G$ on 
translates of $\N$.
\mabsatz
\noindent
{\bf 2.1 THEOREM:} 
\sabsatz
{\it  Let $(\A,\alpha_t)$ be a $C^*$-dynamical system and ${\cal B}$ a 
subalgebra 
such that $\alpha_t {\cal B}\subset {\cal B}$ for $t\geq 0$ and $\cup_t 
\alpha_t 
{\cal B}$ is norm dense in ${\cal A}$. In the GNS representation defined by a 
KMS 
state on ${\cal A}$ at inverse temperature $\beta$ let $\M$ and $\N$ denote the 
weak closures of $\pi({\cal A})$ and
$\pi({\cal B})$ respectively, and $T(t)=\exp(\I tH)$ the unitary group 
implementing 
$\alpha_t$. Denote ${\rm ad}\, T(t) \N=\N(t)$.
Then
\sabsatz
\noindent
(i) The translations $T(t)$ and
the modular group $\Delta_{\N}^{\I u}$, defined by $\N$ and the KMS state 
vector, fulfill the relation (2.20). We have
$${\rm ad}\,\Delta_{\N}^{\I u}\N(t)=\N(\varphi(u,t))\eqno(2.39)$$
with
$$\varphi(u,t)= \frac\beta{2\pi}\log\left\{1+\E^{-2\pi u}
(\E^{{2\pi
t}/\beta}-1)
\right\}\eqno(2.40)$$
for all $u$, $t$ satisfying
$$1+\E^{-2\pi u}
(\E^{{2\pi
t}/\beta}-1)>0.\eqno(2.41)$$
In particular,
$${\rm ad}\,\Delta_{\N}^{\I u}\M\subset\M\eqno(2.42)$$
for all $u\geq 0$, and
$$\N=\bigcap_{u\geq 0}{\rm ad}\,\Delta_{\N}^{\I u}\M.\eqno(2.43)$$
\sabsatz
\noindent 
(ii) The operator $G=\beta H+\log\Delta_{\N}$ is non-negative  and 
essentially self adjoint on a 
common core of $H$ and $\log\Delta_{N}$.  The one 
parameter group $\Gamma(\tau)=\exp(\I\tau G/\beta)$ is given by 
(2.29) and the groups $\Gamma(\tau)$ and $T(t)$ satisfy (2.31). We have
$${\rm ad}\,\Gamma(\tau)\N(t)=\N(\psi(u,t))\eqno(2.44)$$
with
$$\psi(\tau,t)= t+{\beta\over 2\pi}\log
\left\{1+{2\pi\tau\over\beta}\E^{{-2\pi 
t/\beta}}\right\}\eqno(2.45)$$
for all $\tau$, $t$ satisfying 
$$1+{2\pi\tau\over\beta}\E^{{-2\pi 
t/\beta}}>0.\eqno(2.46)$$
In particular,  
$${\rm 
ad}\,\Gamma(\tau)\M\subset \M\qquad{\rm and}\qquad {\rm 
ad}\,\Gamma(\tau)\N\subset \N\eqno(2.47)$$ 
for $\tau\geq 0$, and
$$\N={\rm ad}\,\Gamma(\beta/2\pi)\M.\eqno(2.48)$$ 
\sabsatz
\noindent 
}
{\it Proof:\/} As already noted, the key relations (2.20), (2.29) and (2.31),  
and the 
self adjointness and positivity of $G$  
are a rigorous consequence of the Theorems in [Bch95], [Wie92], [Wie93], 
and [Wie97]. Eqs.\ (2.39) and (2.44) follow directly from (2.20) and (2.29) 
and 
the fact that $\varphi(-u,\varphi(u,t))=t$, 
$\psi(-\tau,\psi(\tau,t))=t$ for $(u,t)$ and $(\tau,t)$ 
satisfying (2.41) and (2.46) respectively. Eqs.\ (2.42), (2.43), (2.47) 
and (2.48) are 
simple consequences of (2.39) and (2.44) since $\cup_{t}{\rm ad}\,T(t)\N$ is 
dense in $\M$.
\bigskip

As a last topic in this section we discuss the relation between the 
translation group $T(t)$ and the modular group $\Delta_\N^{\I u}$. Since 
$T(-\beta u)=\Delta_\M^{\I u}$, one may expect that the actions of $T(-\beta 
u)$ 
and $\Delta_\N^{\I u}$ approximately coincide on elements that have been 
translated far into $\N$, so that \lq\lq boundary effects\rq\rq\ are 
negligible.
That this intuition is indeed solidly founded is the content 
of the next 
two theorems. The first concerns certain matrix elements of the unitary groups,
and gives an estimate for the rate of the convergence. 
The second is about strong convergence of Hilbert space vectors and operators, 
but the error estimates are less explicit.
\mabsatz
{\bf 2.2 THEOREM:} 
\sabsatz
{\it If $A\in \N(t)$ and $B\in \N'$ the following estimate 
holds for  $t>0$ and all $u$:
$$|(B\Omega,\Delta_\N^{\I u }A\Omega)-
(B\Omega,T(-\beta u)A\Omega)|\leq 
2M\,\min\left\{\frac{|\exp (2\pi u)-1|}{\exp(2 \pi 
t/\beta)-1}\,,\,1\right\}\eqno(2.49)$$
with
$$M={\rm max}\,\{\norm{A\Omega}\norm{B\Omega},\norm{A^*\Omega}
\norm{B^*\Omega}\}.\eqno(2.50)$$
}
\medskip
{\it Proof:\/} Consider the two functions
$$
F^+(u)=(B\Omega,\Delta_\N^{-\I u}T( -\beta u) A\Omega),\qquad{\rm 
and}\qquad
F^-(u)=(A^*\Omega,T(\beta u)\Delta_\N^{\I u}B^*\Omega).\eqno(2.51)$$
Theorem A in [Bch95] implies that 
$\Delta_\N^{-\I u}T(-\beta u)$ has a bounded
analytic continuation into the strip $S(-\frac12,0)$. It follows that 
$F^+$ has an analytic continuation into $S(-\frac12,0)$, and $F^-$ 
into $S(0,\frac 12)$. Moreover, by continuity of the unitary groups, 
$F^{\pm}$ is continuous on the real axis.

Denoting 
$j_{\M}={\rm ad}\, J_{\M}$, $j_{\N}={\rm ad}\, J_{\N}$ we obtain
$$\eqalign{
F^+\left(u-\frac\I2\right)&=(B\Omega,J_\N\Delta_\N^{-\I u}T(-\beta u)J_\M 
A\Omega)=
(\Delta_\N^{-\I u}T(-\beta u)j_\M(A)\Omega,j_\N(B)\Omega),\cr
F^-\left(u+\frac\I2\right)&=(A^*\Omega,J_\M T(\beta u)\Delta_\N^{\I u}J_\N 
B^*\Omega)=
(T(\beta u)\Delta_\N^{\I u}j_\N(B^*)\Omega,j_\M(A^*)\Omega).\cr}$$
In particular $F^{\pm}$ is continuous at $u\pm\I/2$, $u\in\R$, and  
$j_\N(B^*)\in\N$ and $j_\M(A^*)\in\M'$ implies
$$F^+\left(u-\frac\I2\right)=F^-\left(u+\frac\I2\right).\eqno(2.52)$$
Moreover, since $T(-s)AT(s)\in \N$ for $s<t$,  we have
$$F^+(u)=F^-(u)\quad {\rm for}\quad u<{t}/\beta.\eqno(2.53)$$
Hence $F^{+}$ and $F^{-}$ have a common analytic continuation to 
a periodic function, $F$, with the period $\I$ and cuts
$[{t}/\beta,\infty)+\I n$, $n\in{\bf Z}$.
This function is  majorized by
$$M={\rm max}\,\{\norm{A\Omega}\norm{B\Omega},\norm{A^*\Omega}
\norm{B^*\Omega}\}.\eqno(2.54)$$

The function $F(z)-F(0)$ vanishes at $z=\I n$, $n\in{\bf Z}$, and is bounded by 
$2M$. 
Therefore, 
$$G(z)=\frac{F(z)-F(0)}{\exp (2\pi z)-1}$$
is analytic and bounded in the same domain as $F$. Along the cuts we 
have 
$|G(z)|\leq 2M(\exp(2\pi t/\beta)-1)^{-1}$. By the  maximum modulus
principle this estimate holds everywhere and thus
$$|(B\Omega,\Delta_\N^{-\I u}T( -\beta u) A\Omega)-
(B\Omega,A\Omega)|\leq 
2M{\frac{|\exp (2\pi u)-1|}{\exp(2 \pi t/\beta)-1}}.\eqno(2.55)$$
This estimate blows up for $t\to 0$, but the left-hand side is 
trivially bounded by $2M$ for all real $u$ and $t$.
Replacing $B\in \N'$ by ${\rm ad}\,\Delta_\N^{-\I u}B\in \N'$
does not change $M$,
so (2.55) gives the desired estimate (2.49).
\vfill\eject
\medskip
\noindent
{\bf 2.3 THEOREM:} 
\sabsatz
\noindent
(i) For every $A\in \M$ and Hilbert space vector $\Psi$
$$\lim_{t\to\infty}\Vert \Delta_\N^{\I u}A(t)\Psi-T(-\beta 
u)A(t)\Psi\Vert=0\eqno(2.56)$$
with $A(t)={\rm ad}\,T(t)A$. The convergence is uniform on half sided 
$u$-intervals $I=(-\infty, u_0]$, $u_0<\infty$.
\sabsatz
\noindent
(ii) For every $A$ in a dense subalgebra of $\M$
$$\lim_{t\to\infty}\Vert {\rm ad}\,\Delta_\N^{\I u}A(t)-{\rm ad}\,T(-\beta 
u)A(t)\Vert=0\eqno(2.57)$$
with uniform convergence on half sided $u$-intervals.
\sabsatz
\noindent
{\it Proof:\/} From (2.48), (2.2) and (2.31) it follows that
$$\Delta_\N^{\I u}=\Gamma(\beta/2\pi)T(-\beta u)\Gamma(-\beta/2\pi)
=T(-\beta u)\Gamma\big((\exp(2\pi u)-1)\beta/2\pi\big).\eqno(2.68)$$
Hence, using (2.31) again,
$${\rm ad}\,\Delta_\N^{\I u}A(t)-{\rm ad}\,T(-\beta u)A(t)
={\rm ad}\,T(t-\beta u)\Big [{\rm ad}\,\Gamma\big(\exp(-2\pi 
t/\beta)h(u)\big)A-
A\Big ]\eqno(2.69)$$
with $h(u)=(\exp(2\pi u)-1)\beta/2\pi$. Now (i) follows from the strong 
convergence of $\Gamma(\tau)$ to $1$ as $\tau\to 0$, because 
$\sup_{u\in I}\vert h(u)\vert <\infty$ for $I=(-\infty,u_0]$.

For general $A\in \M$, $\Vert{\rm ad}\,\Gamma(\tau)A-A\Vert$ need not converge 
to zero as $\tau\to 0$. However, on elements of the form 
$A_g=\int g(\tau) {\rm ad}\,\Gamma(\tau)A\, d\tau$ with $g$ continuous of 
compact 
support, this convergence holds. Moreover, if $g$ is continuously 
differentiable,
then (2.69) implies
$$\Vert {\rm ad}\,\Delta_\N^{\I u}A_g(t)-{\rm ad}\,T(-\beta 
u)A_g(t)\Vert\leq \Vert A\Vert\cdot \Vert dg/d\tau\Vert_1\cdot \sup_{u\in 
I}\vert h(u)\vert\cdot e^{-2\pi t/\beta}.\eqno(2.70)$$

By (2.44) such regularized elements are dense in $\M$ 
if 
the support of $g$ is sufficiently small, and $A_g\to A$ weakly if $g$ tends to 
a delta function.


\babsatz

\noindent
{\bbf 3.\ Two dimensional models}
\mabsatz

The general results of the preceding setting were formulated for a  
$C^*$-dynamical system $(\A, \alpha_t)$ and a subalgebra ${\cal B}$, invariant 
under half-sided shifts by $\alpha_t$. We shall now be more specific and 
consider a  quasi-local algebra $\A$ generated by a local net $\O\mapsto 
\A(\O)$ 
of $C^*$-algebras and 
${\cal B}=\A(\O_0)$ with $\O_0$ a domain invariant under half-sided 
translations 
in the $t$-direction. In the representation $\pi$ generated by a KMS state 
$\omega$ we denote $\pi(\A(\O))^{\prime \prime}$ by $\M(\O)$ and 
$\pi(\A)^{\prime 
\prime}$ by $\M$, as before.

Eqs.\ (2.39) and (2.44) describe the action of the modular- and $\Gamma$-groups 
associated with $\N=\M(\O_0)$ on the translated algebras $\M(\O_0+t{\bf e})$, 
where ${\bf e}$ is the unit vector in the $t$-direction.
We now want to investigate how the groups associated with $M(\O_0)$ 
act on the algebras of more general domains than  $\O_0+t{\bf e}$, in
particular $\O_0+{\bf x}$ with ${\bf x}$ an arbitrary vector in space-time. 
While a general answer to this question appears difficult, the previous results 
lead directly to a description of the action in the case of two 
dimensional theories that factorize in the light cone variables. 

We start by considering local nets in two dimensional 
space time depending only on one light cone variable, i.e.\ nets on a 
light ray.  With $x^0$ the time and $x^1$ the space coordinate of ${\bf 
x}\in\R^2$ the light 
cone variables are $x^{\rm R}=x^0+x^1$ and $x^{\rm L}=x^0-x^1$.  We 
consider either one of them and denote it simply by $x$. 
Note that translations in time $t=x^0$ are equivalent to translations in $x$. A 
local algebra corresponding to an $x$-interval $I\subset \R$ is denoted by 
$\M(I)$. Local commutativity means that $\M(I_1)$ and $\M(I_2)$ commute if 
$I_1\cap I_2=\emptyset$.

We denote the 
modular group for the algebra $\M(\R_+)$ by
$\Delta_+^{\I u}$ and the corresponding group with the positive generator 
$G_+/\beta=H+(1/\beta)\log \Delta_{+}$ by $\Gamma_+(\tau)$. 
We shall also consider the algebra of the negative half axis, $\M(\R_{-})$,
with modular group $\Delta_-^{\I u}$ and the positive operator 
$G_{-}/\beta=H+(1/\beta)\log 
\Delta_{-}$, which generates the group $\Gamma_{-}(\tau)=\exp(\I \tau 
G_{-}/\beta)$. Note that ad $\Gamma_{-}(\tau)$ maps $\M(\R_{-})$ into 
itself for $\tau\leq 0$.

By Eq.\ (2.39)
we have 
$${\rm ad\,}\Delta_+^{\I u}\M([x,\infty[\,)=
\M([\varphi_{+}(u,x),\infty[\,)\eqno(3.1)$$
with
$$\varphi_{+}(u,x)= \frac\beta{2\pi}\log\left\{1+\E^{-2\pi u}
(\E^{{2\pi
x}/\beta}-1)
\right\},\eqno(3.2)$$
for all $x, u\in \R$  such that
$$1+\exp(-2\pi u)[\exp(2\pi 
x/\beta)-1]>0.\eqno(3.3)$$

Note that (3.2) is just the function (2.40). We denote it here by $\varphi_+$ 
because
there is an analogous result for $\M(\R_{-})$:
$${\rm ad\,}\Delta_-^{\I u}\M(\,]-\infty,x]\,)=
\M(\,]-\infty,\varphi_{-}(u,x)]\,)\eqno(3.4)$$
with
$$\varphi_{-}(u,x)=-\varphi_{+}(-u,-x)\eqno(3.5)$$
for  
$$1+\exp(2\pi u)[\exp(-2\pi x/\beta)-1]>0.\eqno(3.6)$$

Likewise, from Eq.\ (2.44)
$${\rm ad\,}\Gamma_{+}(\tau)\M([x_,\infty[\,)=
\M([\psi_{+}(\tau,x),\infty[\,)\eqno(3.7)$$
with
$$\psi_{+}(\tau,x)= x+{\beta\over 2\pi}\log
\left\{1+{2\pi\tau\over\beta}\E^{{-2\pi 
x/\beta}}\right\}\eqno(3.8)$$
for
$$1+(2\pi \tau/\beta)\exp(-2\pi x/\beta)>0,\eqno(3.9)$$
and
$${\rm ad\,}\Gamma_{-}(\tau)\M(\,]-\infty,x]\,)=
\M(\,]-\infty,\psi_{-}(u,x)]\,)\eqno(3.10)$$
with
$$\psi_{-}(\tau,x)= -\psi_{+}(-\tau,-x)\eqno(3.11)$$
for 
$$1-(2\pi \tau/\beta)\exp(2\pi x/\beta)>0.$$
\medskip

We now turn to models in two space-time dimensions 
which can be written 
as a tensor product of one-dimensional models in the light cone variables, 
$x^{\rm R}=x^0+x^1$ and $x^{\rm L}=x^0-x^1$.
For 
a domain $I_{\rm L}\times I_{\rm R}\subset \R^2$ with   $I_{\rm L}$ and 
$I_{\rm R}$ intervals on the $x_{\rm L}$ 
and $x_{\rm R}$ axis respectively, the local algebra is thus
$$\M(I_{\rm L}\times I_{\rm R})=\M(I_{\rm L})\otimes 
\M(I_{\rm 
R}).\eqno(3.13)$$
Here $\otimes $ is the von Neumann tensor product, and $I\mapsto\M(I)$ is 
a local net of von Neumann algebras over $\R$. (For simplicity 
of notation we take identical nets on both axis.) In particular, we 
are interested in the algebras of the forward light cone 
$$\M({\rm V}^+)=\M(\R_{+})\otimes \M(\R_{+})\eqno(3.14)$$ 
and the right wedge 
$$\M(W)=\M(\R_{-})\otimes \M(\R_{+}).\eqno(3.15)$$
The modular groups for these algebras and a factorizing KMS state 
$\omega\otimes \omega$, where $\omega$ is a KMS state for the algebra 
on a light ray, are
$$\Delta_{{\rm V}^+}^{\I u}=\Delta_{+}^{\I u}\otimes \Delta_{+}^{\I 
u}\eqno(3.16)$$
and 
$$\Delta_{W}^{\I u}=\Delta_{-}^{\I u}\otimes \Delta_{+}^{\I 
u}.\eqno(3.17)$$
If ${\bf x}\in\R^2$ we denote the translated 
light cone ${\rm V}^{+}+{{\bf x}}$ by ${\rm V}^+_{{\bf x}}$ and the translated 
wedge $W+{\bf x}$ by $W_{{\bf x}}$. From Eqs.\ (3.1) 
and (3.5) we obtain

\sabsatz
{\bf 3.1 THEOREM:}\sabsatz{\it
$${\rm ad\,}\Delta_{{\rm V}^+}^{\I u}\M({\rm V}^+_{{\bf x}})=
\M({\rm V}^+_{\varphi_{{\rm V}^{+}}(u,{\bf x})})\eqno(3.18)$$
with
$$\varphi_{{\rm V}^{+}}(u,{\bf x})=(\varphi_{+}(u,x^{\rm 
L}),\varphi_{+}(u,x^{\rm 
R}))\eqno(3.19)$$
for $u\in\R$ and ${\bf x}\in \R^2$ such that (3.3) holds for $x=x^{\rm 
L}$ and $x=x^{\rm R}$. 
If $u\geq 0$, then ${\rm ad\,}\Delta_{{\rm V}^+}^{\I u}\M({\rm V}^+_{{\bf 
x}})\subset \M$ for all ${\bf x}\in\R^2$, and if ${\bf x}\in{\rm V}^+$, then
${\rm ad\,}\Delta_{{\rm V}^+}^{\I u}\M({\rm V}^+_{{\bf x}})\subset \M({\rm 
V}^+)$
for all $u$.

Likewise, 
$${\rm ad\,}\Delta_{W}^{\I u}\M(W_{{\bf x}})=
\M(W_{\varphi_{W}(u,{\bf x})})\eqno(3.20)$$
with
$$\varphi_{W}(u,{\bf x})=(\varphi_{-}(u,x^{\rm L}),\varphi_{+}(u,x^{\rm 
R}))\eqno(3.21)$$
for $u\in\R$ and ${\bf x}\in \R^2$ such that (3.3) holds for $x=x^{\rm 
R}$ and (3.6) for $x=x^{\rm L}$. If ${\bf x}\in{W}$, then
${\rm ad\,}\Delta_{{W}}^{\I u}\M({W}_{{\bf x}})\subset \M({
W})$
for all $u$.}
\mabsatz

The flow lines of $\varphi_{{\rm V}^{+}}$ and $\varphi_{W}$ within the 
respective domains are shown 
in Figs.\ 1--2.
\medskip
It is evident from the figures that the character of the modular flow
depends of the distance from the boundary of
the domain considered (forward light cone or wedge). The natural unit 
of length is here the reciprocal temperature, $\beta$. Consider first
the modular group of the forward ligh cone ${\rm V}^+$. 
In terms of the original space 
time coordinates $x^0=(x^{\rm R}+x^{\rm L})/2$ and 
$x^1=(x^{\rm R}-x^{\rm L})/2$ the map (3.19) takes
$(x^0,x^1)$ to $(x^{\prime 0},x^{\prime 1})$ with
$$x^{\prime 0}=x^0-\beta u+R_{{\rm V}^+}^0(x,u),\qquad
x^{\prime 1}=x^1+R_{{\rm V}^+}^1(x,u),\eqno(3.22)$$
where
$$R_{{\rm V}^+}^0(x,u)=
(\beta/4\pi)\log\left\{(1+e^{-2\pi(x^{\rm R}-\beta u)/\beta}
-
e^{-2\pi x^{\rm R}/\beta})(1+e^{-2\pi(x^{\rm L}-\beta u)/\beta}
-
e^{-2\pi x^{\rm L}/\beta})
\right\}\eqno(3.23)$$
and
$$R_{{\rm V}^+}^1(x,u)=
(\beta/4\pi)\log\left\{{1+e^{-2\pi(x^{\rm R}-\beta u)/\beta}
-
e^{-2\pi x^{\rm R}/\beta}\over1+e^{-2\pi(x^{\rm L}-\beta u)/\beta}
-
e^{-2\pi x^{\rm L}/\beta}}
\right\}.\eqno(3.24)$$
{F}ar from the
domain boundary, i.e., for  $x^{\rm R}$, $x^{\rm L}$, $x^{\rm R}- \beta 
u$
and $x^{\rm
L}- \beta u$ large compared to $\beta$, the terms 
$R_{{\rm V}^+}^0$ and
$R_{{\rm V}^+}^1$ are exponentially small, and $\psi_{{\rm 
V}^+}(\cdot,u)$
essentially the same as translation in time by $-\beta u$ in accord 
with Theorems 2.2 and 2.3. 
On the other hand, close to the apex of the light cone (compared to
$\beta$), the action is essentially the same as for $\beta=\infty$,
i.e., dilation by the factor $\exp(-2\pi u)$. 
The deviation from a dilation is
of the order $(|x|/\beta)^2$.

{F}or the wedge $W$ the formulas corresponding to (3.22)--(3.24) are
$$x^{\prime 0}=x^0-\beta u+R_{W}^0(x,u),\qquad
x^{\prime 1}=x^1+R_{W}^1(x,u)\eqno(3.25)$$
with
$$R_{W}^0(x,u)=
(\beta/4\pi)\log
\left\{{1+e^{-2\pi(x^{\rm R}-\beta u)}-
      e^{-2\pi x^{\rm R}/\beta}\over 1+e^{2\pi(x^{\rm L}-\beta u)/\beta}
      -
      e^{2\pi x^{\rm L}/\beta}}
    \right\}\eqno(3.26)$$
and  
$$R_{W}^1(x,u)=
(\beta/4\pi)\log\left\{(1+e^{-2\pi(x^{\rm R}-\beta u)}
    -
    e^{-2\pi x^{\rm R}/\beta})(1+e^{2\pi(x^{\rm L}-\beta u)/\beta}
    -
    e^{2\pi x^{\rm L}/\beta})
    \right\}.\eqno(3.27)$$
Note that the wedge is characterized by $x^{\rm R}\geq 0$ and
$x^{\rm L}\leq 0$. Again the modular action coincides essentially
with time translations far from the domain boundary. Near the edge of the
wedge the coordinate $x^{\rm R}$ is scaled by $\exp(-2\pi u)$ and
  $x^{\rm L}$ is scaled by $\exp(2\pi u)$, up to terms of order
  $(|x|/\beta)^2$. This corresponds to a Lorentz boost, i.e.\ the modular
action at temperature zero.

{F}rom $\Gamma_{\pm}(\tau)$ we can form the one parameter unitary groups
$$\tau\mapsto \Gamma_{\pm}(\tau)\otimes 
\Gamma_{\pm}(\tau)\eqno(3.28)$$
on the tensor product Hilbert space. 
These groups have the positive generators 
$H+(1/\beta)\log\Delta_{\pm,\pm}$, where $\Delta_{\pm,\pm}
=\Delta_{\pm}\otimes 1+1\otimes\Delta_{\pm}$ is the modular operator 
of $\M(\R_{\pm}\times \R_{\pm})$. They correspond respectively to 
the forward and backward light cone ($++$ and $--$) and the left and 
the right wedge ($+-$ and $-+$). All four grops converge to the time 
translations as $\beta\to\infty$. 

The group associated with the forward light cone is
$$\Gamma_{{\rm V}^+}(\tau)=\Gamma_{+}(\tau)
\otimes \Gamma_{+}(\tau).\eqno(3.29)$$
By  (2.44) we have

\mabsatz
{\bf 3.2 THEOREM:}\sabsatz{\it
If ${\bf x}\in \R^2$ and 
$$\tau>-\beta(2\pi)^{-1}\min\{e^{2\pi x^{\rm 
L}/\beta},e^{2\pi x^{\rm 
R}/\beta}\},\eqno(3.30)$$
then
$${\rm ad\,}\Gamma_{{\rm V}^+}(\tau)\M({\rm V}^+_{{\bf x}})=
\M({\rm V}^+_{\psi_{{\rm V}^{+}}(\tau,{\bf x})})\eqno(3.31)$$
with
$$\psi_{{\rm V}^{+}}(\tau,{\bf x})=(\psi_{+}(\tau,x^{\rm 
L}),\psi_{+}(\tau,x^{\rm 
R})).\eqno(3.32)$$
If 
$$\tau>-\beta(2\pi)^{-1}(\min\{e^{2\pi x^{\rm 
L}/\beta},e^{2\pi x^{\rm 
R}/\beta}\}-1)\eqno(3.33)$$ then 
$${\rm ad\,}\Gamma_{{\rm V}^+}(\tau)\M({\rm V}^+_{{\bf x}})\subset
\M({\rm V}^+).\eqno(3.34)$$}
\mabsatz


The group associated with the right wedge,
$$\Gamma_{W}(\tau)=\Gamma_{-}(\tau)\otimes 
\Gamma_{+}(\tau),\eqno(3.35)$$ 
does not induce half sided translations on the wedge algebra, but it 
nevertheless 
acts geometrically for a restricted parameter range. In fact, 
by Eqs. (3.7) and (3.10) we have
\mabsatz
{\bf 3.3 THEOREM:}\sabsatz{\it
If ${\bf x}\in \R^2$ and  
$$-\beta(2\pi)^{-1}e^{2\pi x^{\rm 
R}/\beta}<\tau<\beta(2\pi)^{-1}e^{-2\pi x^{\rm L}/\beta},\eqno(3.36)$$ 
then
$${\rm ad\,}\Gamma_{W}(\tau)\M(W_{{\bf x}})=
\M(W_{\psi_{W}(\tau,{\bf x})})\eqno(3.37)$$
with
$$\psi_{W}(\tau,{\bf x})=(\psi_{-}(\tau,x^{\rm 
L}),\psi_{+}(\tau,x^{\rm 
R})).\eqno(3.38)$$
If 
$$-\beta(2\pi)^{-1}(e^{2\pi x^{\rm 
R}/\beta}-1)<\tau<\beta(2\pi)^{-1}(e^{-2\pi x^{\rm 
L}/\beta}-1),\eqno(3.39)$$
then 
$${\rm ad\,}\Gamma_{W}(\tau)\M(W_{{\bf x}})\subset \M(W).\eqno(3.40)$$
}

The flows of $\psi_{{\rm V}^+}$ and  
$\psi_{W}$ are shown in Figs.\ 3 and 4. 
The groups  $\Gamma_{{\rm V}^+}(\tau)$
and $\Gamma_{W}(\tau)$, approximate the time translations close to the
tip of the light cone and the edge of the wedge, respectively. Indeed,
$\psi_{{\rm V}^+}$ maps $(x^0,x^1)$ to $(x^{\prime 0},x^{\prime 1})$
with
$$x^{\prime 0}=x^0+\tau[\exp(-2\pi x^{\rm R})+\exp(-2\pi x^{\rm 
L})]/2+O(\tau^2/\beta)\eqno(3.41)$$
and
$$x^{\prime 1}=x^1+\tau[\exp(-2\pi x^{\rm R})-\exp(-2\pi x^{\rm 
L})]/2+O(\tau^2/\beta).\eqno(3.42)$$
For $x^{\rm R}$ and $x^{\rm 
L}$ both close to zero, this is close to  $x^{\prime 0}= x^0+\tau$, 
$x^{\prime 1}=x^1$.
More interesting, however, is the 
behavior of $\Gamma_{{\rm V}^+}(\tau)$ far
from the apex of the cone. From Fig.\ 3 one sees clearly that the flow
corresponds to a decelerated motion towards the origin of space. 
More quantitatively, the velocity $v=dx^{\prime 1}/dx^{\prime 0}$
is
$$v=-\tanh (2\pi x^{\prime 1}/\beta)\eqno(3.43)$$
and this differential equation has the general solution
$$x^{\prime 0}(x^{\prime 1})=-(\beta/2\pi)\log(\sinh(2\pi x^{\prime 
1}/\beta))+C\eqno(3.44)$$
where $C$ is an arbitary constant. The path through the origin, $x^{\prime 
1}=0$, corresponds formally to $C=-\infty$.
The flow pattern is invariant 
under a shift in the time direction, in accord with (2.31).

As already mentioned in the Introduction, this flow brings points which start 
out with the velocity of light at infinity 
gradually to rest. Formally we have the reverse of an Unruh effect, for the 
generator of the flow of the observables is positive with the KMS state vector 
as 
a ground state. Measured in terms of the parameter $\tau$ it takes $\beta/2\pi$ 
$\tau$-units for points to reach the forward light cone from infinity. The 
$\tau$-parameter along the path through the origin is related to the real time 
$t$ by
$$t=(\beta/2\pi)\log(1+2\pi \tau/\beta), \quad {\rm i.e.}\quad 
\tau=(\beta/2\pi)(\exp(2\pi t/\beta)-1). \eqno(3.45)$$
The $\tau$-unit is calibrated in such a way, that the $t$- and $\tau$-scales 
coincide precisely where the path hits the apex of the light cone. According 
to Eq.\ (2.31) a different calibration corresponds simply to a shift of the cone 
in the time-direction. It is clear from (3.45) that the $\tau$-parameter is 
\lq\lq slower\rq\rq\ than $t$, in the sense that $d\tau/dt<1$, for a point on 
the path outside the light cone ($\tau<0$), and \lq\lq faster\rq\rq\ than $t$, 
i.e. $d\tau/dt>1$, inside the light cone ($\tau>0$). 

{}For  $\psi_{W}$ the equations corresponding to (3.41) and (3.42) are 
$$x^{\prime 0}=x^0+\tau[\exp(-2\pi x^{\rm R})+\exp(2\pi x^{\rm 
L})]/2+O(\tau^2/\beta)\eqno(3.46)$$
and
$$x^{\prime 1}=x^1+\tau[\exp(-2\pi x^{\rm R})-\exp(2\pi x^{\rm 
L})]/2+O(\tau^2/\beta),\eqno(3.47)$$
and the velocity is 
$$v=-\tanh (2\pi x^{\prime 0}/\beta).\eqno(3.48)$$
Thus the velocity is small close to the space axis, but approaches 
$\pm 1$ far away from the space axis. The explicit solution of (3.48) is
$$x^{\prime 1}(x^{\prime 0})=-(\beta/2\pi)\log(\cosh(2\pi x^{\prime 
0}/\beta))+C.\eqno(3.49)$$
This flow is invariant under a translation in the $x^1$-direction.
The situation is here different from the light cone since not all paths pass 
through the wedge, and those who do, stay in the wedge only for a finite 
$\tau$-interval, cf.\ Eq.\ (3.39). The group even moves localized observables 
out of the global observable algebra in finite 
\lq\lq$\tau$-time\rq\rq, cf.\ Eq.\ (3.36). The direction of acceleration is 
here in the opposite wedge, whereas in the usual Unruh effect it points in the direction of the wedge. In
this sense we have here also a kind of reverse of the situation in the 
Unruh effect.

For the path passing 
through the origin the $\tau$-parameter is related to $t=x^{0}$ by
$$t={\beta\over 4\pi}\log{1+(2\pi\tau/\beta)\over1-(2\pi\tau/\beta)},
\quad{\rm i.e.}\quad \tau={\beta\over 2\pi}\tanh{2\pi t\over\beta}.
\eqno(3.50)$$
The relation to the proper time $t_{\rm p}$ along the path is
$$\tau={\beta\over 2\pi}\sin{2\pi t_{\rm p}\over\beta},\eqno(3.51)$$
i.e., up to a slight deformation $\tau$ is esentially the proper time. We have
$$d\tau/dt_{\rm p}=\cos{2\pi 
t_{\rm p}\over\beta}=(1-(2\pi\tau/\beta)^2)^{1/2},\eqno(3.52)$$
so \lq\lq$\tau$-time\rq\rq\ is everywhere slower than
$t_{\rm p}$ except at the origin (calibration point), where 
both scales coincide with $t$. A change of scale corresponds to a translation 
of the wedge along the $x^1$-axis because of (2.31).

Above we have described the actions of the modular- and $\Gamma$-groups in terms 
of the space time coordinates $(x^{\rm L}, x^{\rm R})$ and also in terms of
$(x^{0}, x^{1})$. The simplest description is 
obtained in yet another coordinate system, that is related to the others by a 
nonlinear transformation. For $x\in \R$ define
$$\xi_\pm=\pm(\beta/2\pi)(\exp(\pm 2\pi x/\beta)-1).\eqno(3.53)$$
The range of  $\xi_+$ is $]-\beta/2\pi,\infty[$  and the range of $\xi_-$ 
is $]-\infty, \beta/2\pi[$ . With $x=x^{\rm L}$ and $x=x^{\rm R}$ respectively 
we thus obtain the four coordinates, $\xi_+^{\rm L},\, \xi_-^{\rm L},\, 
\xi_+^{\rm R}$ 
and $\xi_-^{\rm R}$. In the case of the forward light cone we pick 
$(\xi_+^{\rm L}, \xi_+^{\rm R})$ and in the case of the right wedge 
$(\xi_-^{\rm L}, \xi_+^{\rm R})$ as a curvelinear coordinate system on 
Minkowski space.
In these coordinates the transformations (3.19), (3.21)
for the groups associated with the forward light cone ${\rm V}^+$ become
$$(\xi_+^{\rm L}, \xi_+^{\rm R})\mapsto e^{-2\pi u}(\xi_+^{\rm L}, \xi_+^{\rm 
R}),\qquad 
(\xi_+^{\rm L}, \xi_+^{\rm R})\mapsto (\xi_+^{\rm L}, \xi_+^{\rm 
R})+\tau(1,1)\eqno(3.54)$$ 
and the corrsponding transformations (3.32) and (3.38) for the right 
wedge $W$ are
$$(\xi_-^{\rm L}, \xi_+^{\rm R})\mapsto(e^{2\pi u}\xi_-^{\rm L}, e^{-2\pi 
u}\xi_+^{\rm R}),\qquad 
(\xi_-^{\rm L}, \xi_+^{\rm R})\mapsto(\xi_-^{\rm L}, \xi_+^{\rm 
R})+\tau(1,1).\eqno(3.55)$$
Analogous formulas hold for the backward cone and the left wedge.
Hence in the $\xi$-coordinates the transformations have exactly the same 
form for all $\beta$, including the vacuum case, $\beta=\infty$.

The four coordinate systems $(\xi_\pm^{\rm L}, \xi_\pm^{\rm R})$ can be put 
together by defining
$$(\tilde \xi^{\rm L}, \tilde \xi^{\rm R})=(\beta/2\pi)(
\epsilon(x^{\rm L})(\exp(\epsilon(x^{\rm L})2\pi x^{\rm L}/\beta)-1,
\epsilon(x^{\rm R})(\exp(\epsilon(x^{\rm R})2\pi x^{\rm 
R}/\beta)-1),\eqno(3.56)$$
with $\epsilon(x)=1$ for $x\geq 0$ and $-1$ for $x<0$. This transformation is 
once continuously differentiable, but second derivatives have a discontinuity on 
the light cone. 
The lines corresponding to the flow of the $\Gamma$-groups of the four domains 
(the forward and backward cones and the two wedges) pass continuously through 
the boundaries between the domains, although the groups 
themselves do {\it not} merge to a single one parameter unitary group on the 
Hilbert space.

The transformation (3.56) is of the form $(x^{\rm L}, x^{\rm 
R})\mapsto (f(x^{\rm L}), f(x^{\rm R}))$ with $f$ an order preserving 
bijective map 
$\R\to \R$. Hence it is a causal transformation on two-dimensional space-time, 
i.e., it takes light cones into light cones. 
Such nonlinear causal maps on Minkowski-space exist only in two space-time 
dimensions.

Finally we remark that all results of this section hold for general 
2D theories, provided the state satisfies a KMS condition with respect to 
both light cone coordinates,  $x^{\rm L}$ and $x^{\rm R}$. For factorizing 
states, this holds automatically as a consequence of the KMS condition with 
respect to the time direction. A general proof of a KMS condition with respect 
to light like translations seems out of reach, however, even if one involves 
the relativistic KMS condition [BB94].
\babsatz


\babsatz

\noindent
{\bbf 4. Explicit realizations of modular groups}
\mabsatz

In this section we compute explicitly the modular and $\Gamma$-groups 
for generalized free fields on a light ray and the corresponding 
tensor product models on $\R^2$. In these examples it is 
possible to discuss the action of the groups on the algebras of double 
cones and not only of translated 
forward cones and wedges.

The Weyl algebra of a generalized free Bose field  on a 
light ray is generated by 
elements $W(f)$, with $f$ a real valued Schwartz test function on $\R$, 
satisfying 
the following relations:
$$W(f)^{*}=W(-f)\eqno(4.1)$$
and 
$$W(f)W(g)=e^{-K(f,g)/2} W(f+g)\eqno(4.2)$$
with
$$K(f,g)=
\int_{-\infty}^{\infty}p\, Q(p^2)\tilde{f}(-p), 
\tilde{g}(p) dp,\eqno(4.3)$$
where $Q(p^2)$ is a non-negative polynomial that characterizes the field 
(see [Y93] ). 
Here $\tilde{f}(p)=(1/2\pi)\int\exp(-\I px)f(x)dx$ is the Fourier 
transform 
of $f$. 
The kernel of $\K$ of $K$, defined by $K(f,g)=\int 
\K(y-x)f(x)g(y)dxdy$,  is
$$\K(y-x)=M(-id/dy)\delta(y-x)\eqno(4.4)$$
with $M(p)=pQ(p^2)$, so $W(f)$ and $W(g)$ commute if $f$ and $g$ have 
disjoint 
supports.  

Translations in time are equivalent to translations along the light ray 
and are represented by automorphisms of 
the Weyl algebra,
$$\alpha_{t}(W(f))=W(f(\cdot-t)).\eqno(4.5)$$

A quasi free KMS state $\omega$ at inverse temperature $\beta$ 
is defined on the Weyl algebra by
$$\omega(W(f))=\exp(-\omega_{2}(f,f)),\eqno(4.6)$$
where $\omega_{2}$ is given by a positive definite kernel 
$\W_2(y-x)$ (two point function) that is analytic in the strip 
$S(0,\beta)$ and satisfies
$$\W_2(\xi)-\W_2(-\xi)=\K(\xi)\eqno(4.7)$$
for real $\xi$, together with the KMS condition
$$\W_2(\xi+\I\beta)=\W_2(-\xi).\eqno(4.8)$$
It is straightforward to show that these conditions fix 
$\W_2$ (up to normalization); its Fourier transform is
$$\tilde\W_2(p)=\frac{pQ(p^2)}{1-\E^{-\beta p}}.\eqno(4.9)$$

The Fourier transform of the meromorphic function $(1-\exp(-\beta 
p))^{-1}$ 
is seen to be $\lim_{\varepsilon\to 
0_{+}}(2\pi 
\I\beta)^{-1}(\exp(\beta^{-1}2\pi \xi+\I\varepsilon)-1)^{-1}$ by contour 
integration.  The 
Fourier transform of (4.9) for general $Q$
follows by differentiation. In particular we have for
$Q\equiv 1$, i.e. a field of scaling dimension $1$,
$$\W_2(\xi)=\lim_{\varepsilon\to 0_{+}}\frac1{\beta^2}
\frac1{\bigl(\sinh\frac{\pi(\xi+\I\varepsilon)}\beta\bigr)^2},\eqno(4.10)
$$
and for $Q$ a polynomial of degree $n$ 
$$\W_2(\xi)=\lim_{\varepsilon\to 0_{+}}
\frac{P(\cosh\frac{\pi \xi}\beta,\sinh\frac{\pi \xi}\beta)}
{\bigl(\sinh\frac{\pi(\xi+\I\varepsilon)}\beta\bigr)^{2n+2}},\eqno(4.11)$$
where $P$ is a polynomial in two variables. We shall restrict 
ourselves to the case that $Q(p^2)=p^{2n}$, i.e.\ a field of a definite 
scaling dimension $(n+1)$, in order not to mix 
the effects coming from the non-zero temperature with those due to 
inhomogeneous polynomials $Q$ (see [Y93] for the latter).

Denoting the Weyl operators corresponding to $Q(p^2)=p^{2n}$ by 
$W^{(n)}(f)$ it is clear from (4.3) that we may identify 
$$W^{(n)}(f)=W^{(0)}(\I^nf^{(n)}),\eqno(4.12)$$ 
where $f^{(n)}$ is the $n$-the derivative of $f$, and from (4.9) we see 
also 
that 
a KMS state for any $n$ is the same as 
the KMS state for $n=0$ restricted to the operators 
$W^{(0)}(\I^nf^{(n)})$.
This will allow us to reduce everything to 
the simplest case, $n=0$. 

Let $\pi$ be the GNS representation defined by the KMS 
state (4.6) on the Weyl algebra of the $W^{(0)}(f)$'s. If $I\subset\R$ is 
an 
interval, bounded or unbounded, 
we define $\M^{(n)}(I)$ to be the von 
Neumann algebra generated by $\pi(W^{(n)}(f))$ with supp $f\subset I$.
Because of the identification discussed above these algebras are for 
all $n$ realized on the same Hilbert space. 

By exactly the same arguments as in [Y93], Sec.\ 3, one proves
\mabsatz

\noindent
{\bf 4.1 LEMMA:}\sabsatz{\it If $I$ is an unbounded interval, then
$\M^{(n)}(I)\equiv \M(I)$ is independent of $n$. If $I$ is bounded
with a non-empty interior, then $\M^{(m)}(I)$ is a proper subalgebra of 
$\M^{(n)}(I)$ for $m>n$.}
\mabsatz 

This lemma implies in particular that the modular 
operator $\Delta_{+}$ corresponding to the half-line $\R_{+}$ is the 
same for all $n$. 

The main result about the modular action is the following:
\mabsatz

\noindent
{\bf 4.2 THEOREM:}\sabsatz{\it Let $\omega$ be the quasi free 
KMS state (4.6) and $\pi$ the corresponding representation of the 
Weyl algebra for $n=0$.
The modular group of $\M(\R_{+})$ defined by 
$\omega$ 
has the form
$$
\Delta_+^{\I u}\pi(W^{(0)}(f))\Delta_+^{-\I u}=
\pi(W^{(0)}(\delta^{(0)}_uf))\eqno(4.13)$$
with
$$\delta^{(0)}_uf(x)=f\left(\frac\beta{2\pi}\log\Bigl\{1+\E^{2\pi u}
(\E^{{2\pi
x}/\beta}-1)
\Bigr\}\right)\eqno(4.14)$$
for supp $f\subset \R_{+}$.}
\mabsatz

{\it Remark 1.\/} It is understood that if supp $f\subset \R_{+}$, 
then 
also $\delta^{(0)}_uf(x)=0$ for all 
$x<0$.

{\it Remark 2.\/} The cyclic vector $\Omega$ corresponding to $\omega$ 
has the Reeh-Schlieder property, in particular it is cyclic for 
$\M(\R_{+})$.  Hence (4.13) with supp $f\subset \R_{+}$, together with 
$\Delta_+^{\I t}\Omega=\Omega$, already fixes $\Delta_+^{\I t}$ as a 
unitary group on the GNS Hilbert space.  But $\Delta_+^{\I 
u}\pi(W^{(0)}(f))\Delta_+^{-\I u}$ is, 
of course, a well defined operator on the Hilbert space for all $f$ of 
compact support, and in fact if $u\geq 0$, then (4.13) and (4.14) 
hold for functions with support outside of $\R_{+}$ with the
understanding that (4.14) is zero when the argument of the logarithm 
is $\leq 0$.   This is a simple 
consequence of (3.1)--(3.3). If $u<0$, however, the transformed operator only 
belongs to the observable algebra if condition (3.3) holds on the support of 
$f$.

\mabsatz

{\it Proof of Theorem 4.2:\/} The formula (4.14) is motivated by 
Eq.\ 
(2.20).
In order to show that it is the correct formula for the modular action 
we have to check the following properties of $\delta^{(0)}_u$:
\sabsatz
\item{(i)} $\delta^{(0)}_u$ maps the space of test functions with 
support in $\R_{+}$ into itself.
\item{(ii)} The group property, i.e. $\delta^{(0)}_u\circ\delta^{0}_{u'}
=\delta_{u+u'}^{(0)}$.
\item{(iii)} The unitarity of $\delta^{(0)}_u$ in the scalar product 
defined by the two point function (4.10).
\item{(iv)} The KMS condition: For real test functions $f$ and $g$ with 
support in $\R_{+}$ the function 
$u\mapsto\omega_{2}(f,\delta^{(0)}_{u}g)$ has 
an analytic continuation into the strip $S(-1,0)$ and
$$
\omega_{2}(f,\delta_{u-\I}^{(0)}g)
=\omega_{2}(\delta_{u}^{(0)}g,f).
$$
\sabsatz 
Property (i) is obvious from the definition, and (ii) and (iii) are 
straightforward calculations.  We now check the KMS condition. Put
$$L(u,x):=\frac\beta{2\pi}\log\left\{1+\E^{2\pi u}
(\E^{{2\pi
x}/\beta}-1)
\right\}.\eqno(4.15)$$
Since $L(u,L(-u,y))=y$ (group property) we have
$$\omega_{2}(f,\delta_{u}^{(0)}g)=\int\int
\W_2(L(-u,y)-x){\partial L(-u,y)\over 
\partial y}\,{f(x)}g(y)dxdy.\eqno(4.16)$$
Using the 
addition formula for hyperbolic functions, we compute for the two 
point function (4.10):
$$\eqalign{\W_2&(L(-u,y)-x)\,{\partial L(-u,y)/\partial 
y}\cr
=\frac1{4\beta^2}&\Bigl[\sinh(\pi L(-u,y)/\beta)\cosh(\pi x/\beta )-
\cosh(\pi L(-u,y)/\beta)\sinh(\pi x/\beta )
+\I\varepsilon\Bigr]^{-2}
\frac{\partial L(-u,y)}{\partial y}\cr
=\frac1{16\beta^2}&\Bigl[\Bigl\{
\left(1+\E^{-2\pi u}(\E^{{2\pi y}/\beta}-1)\right)^{1/2}+
\left(1+\E^{-2\pi u}(\E^{{2\pi y}/\beta}-1)\right)^{-1/2}\Bigr\}
\cosh(\pi x/\beta)\cr
-&\Bigl\{
\left(1+\E^{-2\pi u}(\E^{{2\pi y}/\beta}-1)\right)^{1/2}+
\left(1+\E^{-2\pi u}(\E^{{2\pi y}/\beta}-1)\right)^{-1/2}\Bigr\}
\sinh(\pi x/\beta)+\I 
\varepsilon
\Bigr]^{-2}\times\cr
&\times\frac{\E^{-2\pi u}\E^{{2\pi y}/\beta}}{\E^{-2\pi u}
(\E^{{2\pi y}/\beta}-1)+1}\cr
=\frac1{16\beta^2}&\Bigl[
\bigl\{\E^{-2\pi u}(\E^{{2\pi y}/\beta}-1)\bigr\}\cosh(\pi x/\beta)
-\bigl\{2+
\E^{-2\pi u}(\E^{{2\pi y}/\beta}-1)\bigr\}\sinh(\pi x/\beta)
+\I 
\varepsilon\Bigr]^{-2}\times\cr
&\times
\E^{-2\pi u}\E^{{2\pi y}/\beta}\cr
=\frac{\E^{{2\pi y}/\beta}}{16\beta^2}&\Bigl[
\E^{-\pi u}(\E^{{2\pi y}/\beta}-1)[\cosh(\pi x/\beta)
-\sinh(\pi x/\beta)]
-
\E^{\pi u}2\sinh
(\pi x/\beta)
+\I\varepsilon\Bigr]^{-2}.\cr
}$$
For $x,y>0$, $e^{-\pi u}$ comes with a positive factor and 
 $e^{\pi u}$ with a negative one. For
$\varepsilon>0$,  the total expression is therefore analytic in $u$ in
the strip $S(-1,0)$, and this analyticity is preserved in the limit 
$\varepsilon\to 0_{+}$ after smearing in $x$ and $y$ with test 
functions with support in $\R_{+}$. The boundary value at $u-i$, 
$u\in\R$, is
$$\frac{\E^{{2\pi y}/\beta}}{16\beta^2}\Bigl[
\E^{\pi u}2\sinh
(\pi x/\beta)
-\E^{-\pi u}(\E^{{2\pi y}/\beta}-1)[\cosh(\pi x/\beta)
-\sinh(\pi x/\beta)]+\I\varepsilon
\Bigr]^{-2}.
$$
This is precisely $\W_2(x-L(-u,y))\,{\partial L(-u,y)/\partial 
y}$ (by the same computation). Hence the KMS condition is verified.
\mabsatz

The representation of the group $\Gamma_{+}(\tau)=\exp(\I\tau 
G_{+}/\beta)$ with the positive 
generator $G_{+}/\beta=H+(1/\beta)\log \Delta_{+}$ now follows 
immediately 
from Eqs\ (2.25) and (4.13)--(4.14):
\mabsatz
\noindent
{\bf 4.3 THEOREM:}\sabsatz {\it For $\tau\geq 0$ and all $f$
$$
\Gamma_{+}(\tau)\pi(W^{(0)}(f))\Gamma_{+}(-\tau)=
\pi(W^{(0)}(\gamma_{\tau}^{(0)} f))\eqno(4.17)$$
with 
$$\gamma_{\tau}^{(0)} f(x)=f\left(x+{\beta\over 2\pi}\log
\left\{1-{2\pi\tau\over\beta}\E^{-{2\pi x/\beta}}\right\}\right).
\eqno(4.18)$$
}
\mabsatz
{\it Remark:\/} It is understood that $\gamma_{\tau}^{(0)} f(x)=0$ if 
the argument of the logarithm is $\leq 0$, i.e., if $x\leq 
-\beta/(2\pi)\log(2\pi \tau/\beta)$. Note that if $\tau\geq 
\beta/(2\pi)$, then supp $\gamma_{\tau}^{(0)}f\subset 
\R_{+}$ for any $f$ of compact support.

By (4.12) we obtain as a corollary of Theorems 4.2 and 4.3
\mabsatz

\noindent
{\bf 4.4 THEOREM:}\sabsatz{\it For $n>0$ the action of {\rm 
ad}$\,\Delta_+^{\I 
u}$
and {\rm ad}$\,\Gamma_{+}(\tau)$ on $W^{(n)}(f)$ 
with {\rm supp }$f\subset \R_{+}$
is
$$
\Delta_+^{\I u}\pi(W^{(n)}(f))\Delta_+^{-\I u}=
\pi(W^{(n)}(\delta^{(n)}_uf))\eqno(4.19)$$
with 
$$\delta^{(n)}_uf(x)=\int_{0}^xdx_{1}\int_{0}^{x_{1}}\cdots 
\int_{0}^{x_{n-1}}dx_{n}\delta^{(0)}_uf^{(n)}(x_{n}),\eqno(4.20)$$
and for $\tau\geq 0$
$$
\Gamma_{+}(\tau)\pi(W^{(n)}(f))\Gamma_{+}(-\tau)=
\pi(W^{(n)}(\gamma_{\tau}^{(n)} f))\eqno(4.21)$$
with
$$\gamma^{(n)}_{\tau} f(x)=\int_{0}^xdx_{1}\int_{0}^{x_{1}}\cdots 
\int_{0}^{x_{n-1}}dx_{n}\gamma^{(0)}_{\tau} f^{(n)}(x_{n}).\eqno(4.22)$$
}
\mabsatz

{\it Remark.\/} 
It should be noted that $\delta_{u}^{(n)}f$ is in general no longer a 
test function if $n>0$, for it may behave like $x^{n-1}$ for $x\to 
\infty$. However, it belongs to the Hilbert space defined by the two 
point function and hence the Weyl operators are well defined. The same 
applies to $\gamma_{\tau}^{(n)}f$.
\mabsatz 

Next we investigate the localization properties of the modular 
groups. We recall from Lemma 4.1 that for an unbounded interval 
$[x,\infty[$ the 
algebras $\M^{(n)}([x,\infty[\,)\equiv\M([x,\infty[\,)$ are 
independent of $n$. Hence the general result (3.1) applies. For the 
algebras corresponding to bounded intervals we have

\mabsatz
\noindent
{\bf 4.5 THEOREM:}\sabsatz{\it
For $-\infty<x<y<\infty$ and $u$ and $\tau$ restricted according to (3.3), 
(3.5) (3.9), (3.12) 
$${\rm ad\,}\Delta_+^{\I u}\M^{(0)}([x,y])=
\M^{(0)}([\varphi_{+}(u,x),\varphi_{+}(u,y)]),\eqno(4.23)$$
and 
$${\rm ad\,}\Gamma_{+}(\tau)\M^{(0)}([x,y])=
\M^{(0)}([\psi_{+}(\tau,x),\psi_{+}(\tau,y)]).\eqno(4.24)$$
{}For
$n>0$ a local algebra 
$\M^{(n)}([x,y]))$ is not mapped into 
an $\M^{(n)}(I)$ with bounded $I$.}
\mabsatz
{\it Proof:\/} For fixed $u$ and $\tau$ the maps 
$x\mapsto \varphi_{+}(u,x)$ and  $x\mapsto 
\psi_{+}(\tau,x)$ 
are one to one for $x$ satisfying (3.3) and (3.9) respectively, and the inverse 
maps correspond to $u\to -u$ and $\tau\to -\tau$. 
{}From (3.14) it is clear that $f$ has its 
support in
$[x,y]$, iff $\delta_{u}^{(0)}f$ has its support in
$[\varphi_{+}(u,x),\varphi_{+}(u,y)]$ iff $\gamma_{\tau}^{(0)}f$ 
has its support in
$[\psi_{+}(\tau,x),\psi_{+}(\tau,y)]$.  Hence (4.23) and (4.24) follows 
directly from Theorems 4.2 and 4.3. 

To show the dislocalization for 
$n>0$ we note first that neither
$\delta_{u}^{(0)}f^{(n)}$ nor $\gamma_{\tau}^{(0)}f^{(n)}$  
is a derivative of a function with 
compact support (except for $f\equiv 0$). This is easily seen by 
considering the
Fourier transforms of these functions, divided by $p$;
the $1/p$ singularity is not compensated by the derivatives 
because of the non-linear variable 
transformations, and  analyticity is lost.
Consider now a bounded interval $I$ and a function 
$g$ such that $g^{(n+1)}$ vanishes on $I$. Then $W^{(0)}(g)$ belongs 
to the commutant of $\M^{(n)}(I)$. If 
$W^{(n)}(\delta_{u}^{(n)}f)=W^{(0)}(\delta_{u}^{(0)}f^{(n)})$ 
would belong to  $\M^{(n)}(I)$, then 
it would commute with $W^{(0)}(g)$, which means that
$$\int\delta_{u}^{(0)}f^{(n)}(x)\,g'(x)dx=0.$$
This must in particular hold for all $g$ with $g'\equiv 1$ on $I$ 
because such $g$ fulfill $g^{(n+1)}=0$ on $I$ for $n>0$. Hence
$$\int_{I}\delta_{u}^{(0)}f^{(n)}(x)dx=0.\eqno(4.25)$$ 
By isotony this should also hold for all larger intervals, and hence 
$\delta_{u}^{(0)}f^{(n)}$ would be a derivative of a 
function of compact support. As remarked above, this is not the case, and 
we 
have a contradiction to the assumption that 
$W^{(n)}(\delta_{u}^{(n)}f)$ belongs to $\M^{(n)}(I)$ with $I$ bounded. 
By the same argument $W^{(n)}(\gamma_{\tau}^{(n)}f)$ does not belong to 
$\M^{(n)}(I)$.

\mabsatz
{\it Remark 1.\/} In terms of the field operators $\Phi^{(n)}(x)$, 
defined by 
$$\pi(W^{(n)}(f))=\exp(\I\hbox{$\int$}\Phi^{(n)}(x)f(x)dx),\eqno(4.26)$$ 
Eqs.\ 
(3.13) and (3.17) say that
$$\Delta^{\I u}_{+}\Phi^{(0)}(x)\Delta^{-\I 
u}_{+}=\Phi^{(0)}\left(\varphi_{+}(u,x)\right)
{\partial \varphi_{+}(u,x)\over \partial 
x}\eqno(4.27)$$
and
$$
\Gamma_{+}(\tau)\Phi^{(0)}(x)\Gamma_{+}(-\tau)=
\Phi^{(0)}\left(\psi_{+}(\tau,x)\right){\partial \psi_{+}(\tau,x)\over 
\partial 
x}.\eqno(4.28)
$$
In particular we have 
$$\Delta^{\I u}_{+}\Phi^{(0)}(0)\Delta^{-\I 
u}_{+}=e^{-2\pi u}\Phi^{(0)}(0)\eqno(4.29)$$
and
$$\Gamma_{+}(\tau)\Phi^{(0)}(0)\Gamma_{+}(-\tau)
=(1+(2\pi \tau/\beta))^{-1}\Phi^{(0)}
\left((\beta/2\pi)\log(1+(2\pi \tau/\beta))\right).
\eqno(4.30)$$
(Although the field is only an operator valued distribution, these 
equations have a rigorous meaning in terms of quadratic forms.)
Conversely (3.33) and (3.34), together with Eqs.\ (2.20) and (2.26), 
imply (3.31) and (3.32). For $n>0$, however, ad$\Delta^{\I u}_{+}$ is a 
non-local 
transformation of the field operators by Thm.\ 4.4. For instance we 
have
$$\Delta^{\I u}_{+}\Phi^{(1)}(0)\Delta^{-\I 
u}_{+}=e^{-2\pi u}\Phi^{(1)}(0)-(2\pi/\beta)e^{-4\pi u}
\int_{0}^{\infty}\Phi^{(1)}(x)dx.
\eqno(4.31)$$
This shows clearly that there is more to the transformation law for 
the fields than 
Eqs.\ (2.20) and (2.29) alone.
\mabsatz

If $\M(\R_{+})$ is replaced by $\M(\R_{-})$ the previous results 
apply with
appropriate changes of signs, cf. (3.5). 

Forming tensor product algebras as in (3.13) we obtain generalized 
free fields on two dimensional space-time and KMS states that 
factorize in the light cone variables.
In the case of 
the field with lowest scaling dimension, i.e., $n=0$, the double cone algebras 
$\M^{(0)}(I_{\rm L}\times I_{\rm R})$, with $I_{\rm L}$, $I_{\rm R}$ 
bounded intervals, are again mapped into algebras of double cones. 
The flow lines of Figs.\ 1-4 describe in this case not only the movement of the 
apex 
of a forward light cone or the edge of a wedge, but also the movement 
of the double cones. 

For fields of higher scaling dimension, i.e. $n>0$, 
however, 
double cone algebras are after the transformation no longer localized 
in double cones within the net $\M^{(n)}$.  They are still localized 
in double cones within the net $\M^{(0)}(\cdot)$, because 
$\M^{(n)}(\cdot)$ is a subnet of $\M^{(0)}(\cdot)$.

\babsatz

\babsatz

\noindent
{\bbf 5. Conclusions}
\mabsatz

In a KMS state at inverse temperature $\beta$ the time translations 
coincide (up to a sign and scaling by $\beta$) with the modular group 
of the global observable algebra.  From this fact, and the general 
theory of half-sided modular inclusions, algebraic relations between 
time translations and the modular groups for certain domains of 
space-time can be derived.  The action of the modular groups on 
observables localized inside these domains far from the boundary is 
approximately given by the time translations.  In two dimensional 
models and states that satisfy a KMS condition with respect to 
light-like translations (in particular models that factorize in the 
light cone coordinates), a geometric interpretation can be given of 
the action of the modular groups of the forward light cone and a 
space-like wedge on observable algebras localized in translated 
domains of the same type.  This action can be studied in detail in 
simple free field models.  Besides the modular groups, the theory also 
leads to one parameter groups with positive generators, for which the 
KMS state is a ground state.  The actions of these groups for the 
forward cone and the wedge can also be described geometrically and 
interpreted, at least formally, as a kind of a reverse Unruh effect.

\mabsatz

\bigskip
\noindent
{\bbf Acknowledgements}
\mabsatz
We thank D.\ Buchholz, H. Narnhofer,  P.\ Michor and W. Thirring 
for helpful comments. Hospitality of the  
the Erwin Schr\"odinger Institute, Vienna
(H.J.B.) and the University Science Institute, Reykjavik (J.Y.) is 
also gratefully acknowledged.
%
\bigskip
\bigskip
\noindent
{\bbf References}
\mabsatz
\def\ref{\par\noindent\hangindent=\parindent\ltextindent}
\def\ltextindent#1{\hbox to \hangindent{#1\hss}\ignorespaces}
{\parindent=1.3cm
{\baselineskip=3ex\eightpoint\smallskip
\font\eightit=cmti8
\font\eightbf=cmbx8
\def\it{\eightit}
\def\bf{\eightbf}
\ref {[BB94]} J. Bros, D.\ Buchholz:
{\it Towards a relativistic KMS-condition},
Nucl.\ Phys. {\bf B 429},2911-318 (1994).
\ref {[BW75]} J. Bisognano and E.H. Wichmann:
{\it On the duality condition for a Hermitean scalar field},\newline
J. Math. Phys. {\bf 16}, 985-1007 (1975).
\ref {[BW76]} J. Bisognano and E.H. Wichmann:
{\it On the duality condition for quantum fields},
J. Math. Phys. {\bf 17}, 303-321 (1976).
\ref {[Bch92]} H.-J. Borchers:
{\it The CPT-Theorem in Two-dimensional Theories of Local 
Observables}
Commun. Math. Phys. {\bf 143}, 315-332 (1992).
\ref {[Bch95]} H.-J. Borchers: {\it On the use of modular groups in 
quantum field theory}, Ann.\ Inst.\ H.\ Poincar\'e {\bf 64}, 331-382 
(1996).
\ref {[Bch98]} H.-J. Borchers: {\it Half-sided Translations and the 
Type of von Neumann Algebras}, 
Lett.Math.Phys., to appear (1998)
\ref {[BR79]} O. Bratteli, D.W. Robinson:
{\it Operator Algebras and Quantum Statistical Mechanics I},
Springer Verlag, New York, Heidelberg, Berlin (1979).
\ref {[Bu78]} D. Buchholz: {\it On the Structure of Local Quantum Fields
with Non--trivial Interactions}, In: Proceedings of the International
Conference on Operator Algebras, Ideals and their Applications in
Theoretical Physics, Leipzig 1977, Teubner--Texte zur Mathematik (1978)
p. 146-153.
\ref{[BDL90]} D. Buchholz, C. D'Antoni, R. Longo: {\it
  Nuclear Maps and Modular Structures II: Applications to Quantum Field
Theory}, Commun. Math. Phys. {\bf 129}, 115-138 (1990).
\ref {[ENTS95]} G.G.\ Emch, H.\ Narnhofer, W.\ Thirring, G.L.\ Sewell: 
{\it Anosov actions on noncommutative algebras}, J.\ Math.\ Phys.\ {\bf 
35}, 
5582-5598 (1994)
\ref {[Ha96]} R. Haag: {\it Local Quantum Physics}, 
Springer Verlag, 2nd ed., Berlin-Heidelberg-New York (1996). 
\ref {[HL82]} P.D. Hislop and R. Longo:
{\it Modular structure of the local algebra associated with a free
    massless scalar field theory},
Commun. Math. Phys. {\bf 84}, 71-85 (1982).
\ref {[KR86]} R.V. Kadison and J.R. Ringrose: {\it Fundamentals of the 
Theory
of Operator Algebras} II, New York: Academic press, (1986).
\ref {[RS61]} H. Reeh and S. Schlieder: {\it Eine Bemerkung zur 
Unit\"ar\"aquivalenz von Lorentzinvarianten \newline Feldern}, 
Nuovo Cimento {\bf 22}, 1051 (1961).
\ref{[Sew80]} G.L. Sewell: {\it Relativity of temperature and the 
Hawking effect}, Phys. Lett. {\bf 79A}, 23-24 (1980)
\ref{[Sew82]} G.L. Sewell: {\it Quantum fields on manifolds: PCT and 
gravitationally induced thermal states}, Ann.\ Phys.\ {\bf 141}, 201-224 (1982)
\ref {[Ta70]} M. Takesaki: {\it Tomita's Theory of Modular Hilbert 
Algebras 
and
its Applications}, Lecture Notes in Mathematics, Vol. {\bf 128}
Springer-Verlag Berlin, Heidelberg, New York (1970).
\ref{[U76]} W.G. Unruh:  {\it Notes on 
black-hole evaporation}, Phys. Rev. {\bf D14}, 870-892 (1976) 
\ref {[Wie92]} H.-W. Wiesbrock: {\it A comment on a recent 
work of Borchers},
Lett.Math.Phys. {\bf 25}, 157-159 (1992).
\ref {[Wie93]} H.-W. Wiesbrock: {\it Half-Sided Modular Inclusions of von
Neumann Algebras}, Commun. Math. Phys. {\bf 157}, 83 (1993)
\ref {[Wie97]} H.-W. Wiesbrock: {\it Half-Sided Modular Inclusions of von
  Neumann Algebras}, Erratum,  Commun. Math. Phys. {\bf 184}, 683-685 
(1997)
  \ref {[Y93]} J.\ Yngvason:
{\it A note on Essential Duality},
Lett. Math. Phys. {\bf 31},127-141 (1993)

\vfill\eject
\vbox to\vsize{\vfil
\hbox to\hsize{\hfil{\epsfysize=7cm\epsffile{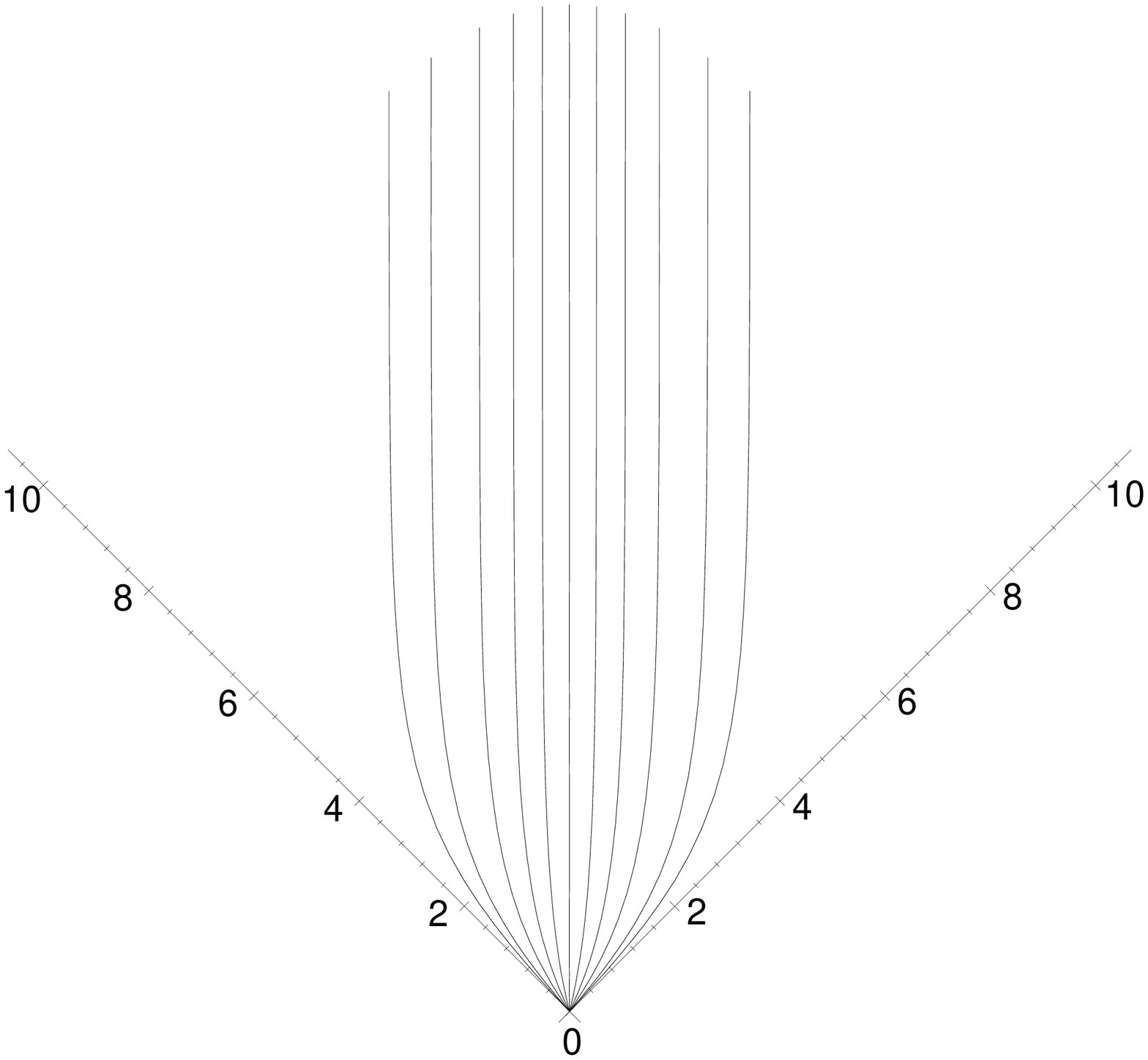}\hfil}}
\vskip\baselineskip
\hbox to\hsize{\hfil Figure 1: The modular flow in the forward light cone. 
The unit is inverse temperature, $\beta$.\hfil}
\vfil
\hbox to\hsize{\hfil{\epsfysize=7cm\epsffile{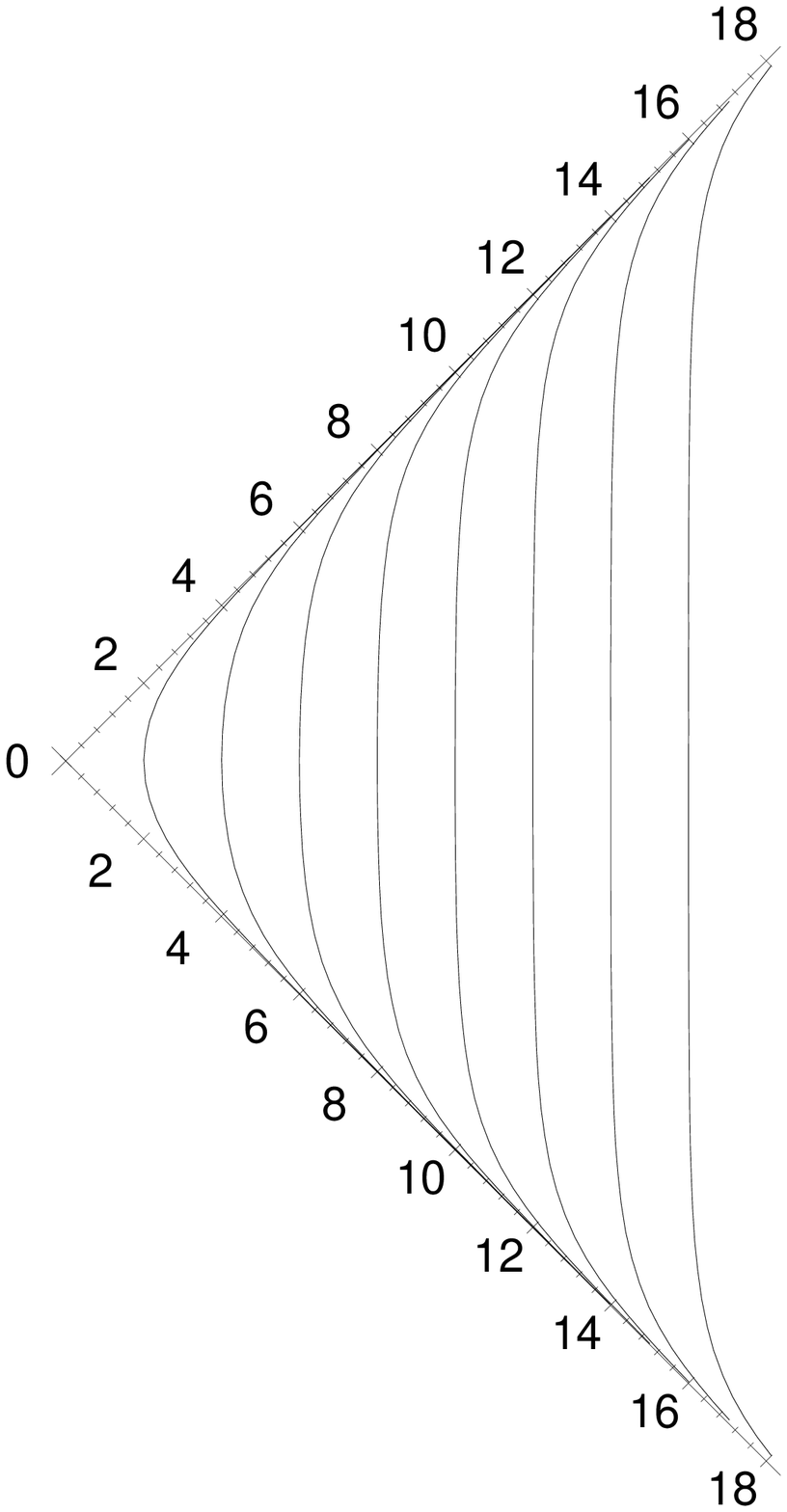}\hfil}}
\vskip\baselineskip
\hbox to\hsize{\hfil Figure 2: The modular flow in a space-like wedge\hfil}
\vfil
}
\vfill\eject
\vbox to\vsize{\vfil
\hbox to\hsize{\hfil{\epsfysize=7cm\epsffile{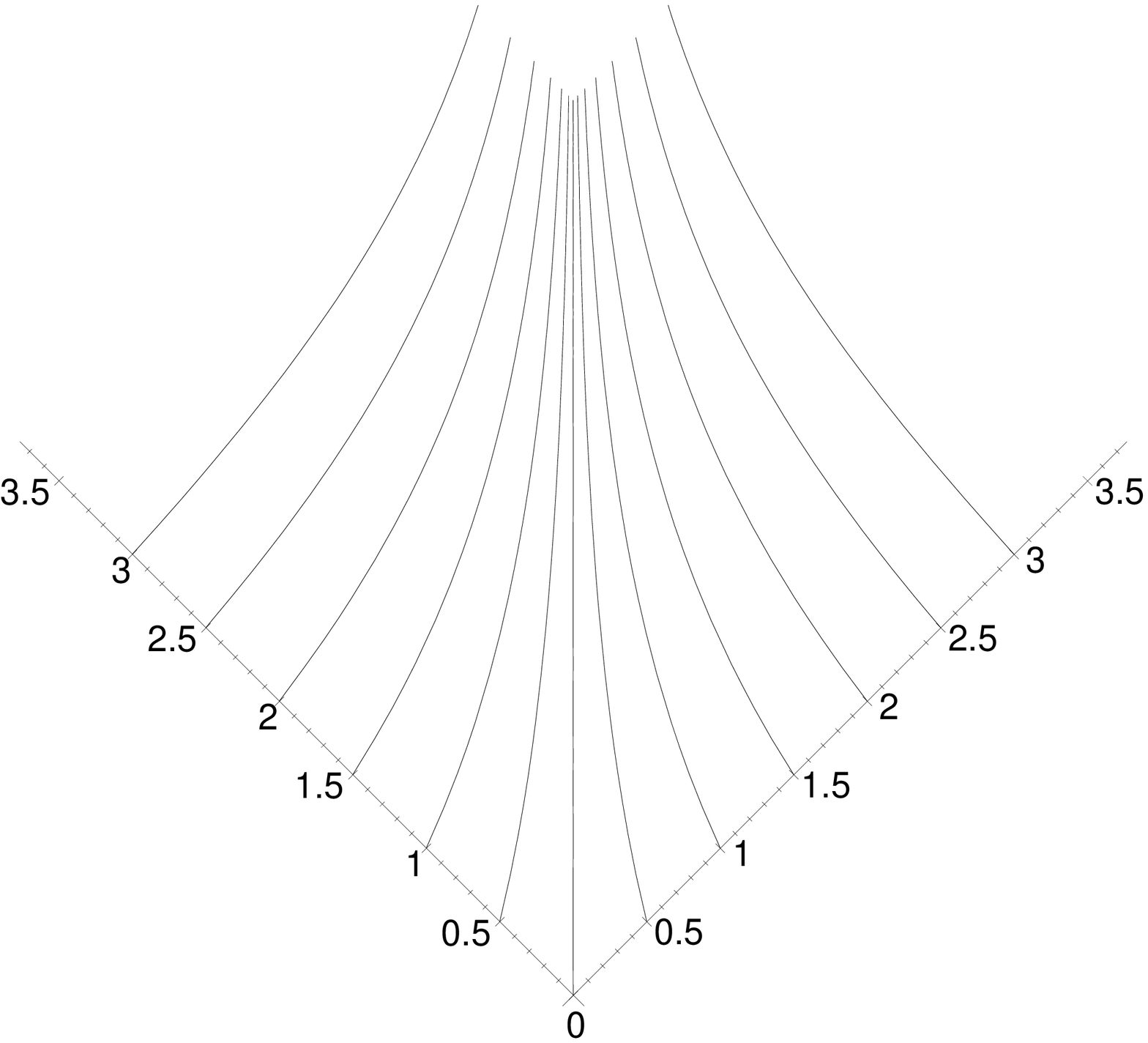}\hfil}}
\vskip\baselineskip
\noindent Figure 3: The flow of $\Gamma_{{\rm V}^+}(\tau)$ within the 
forward light cone. The whole pattern is invariant under translations in the 
$x^{0}$-direction.
\vfil
\hbox to\hsize{\hfil{\epsfysize=7cm\epsffile{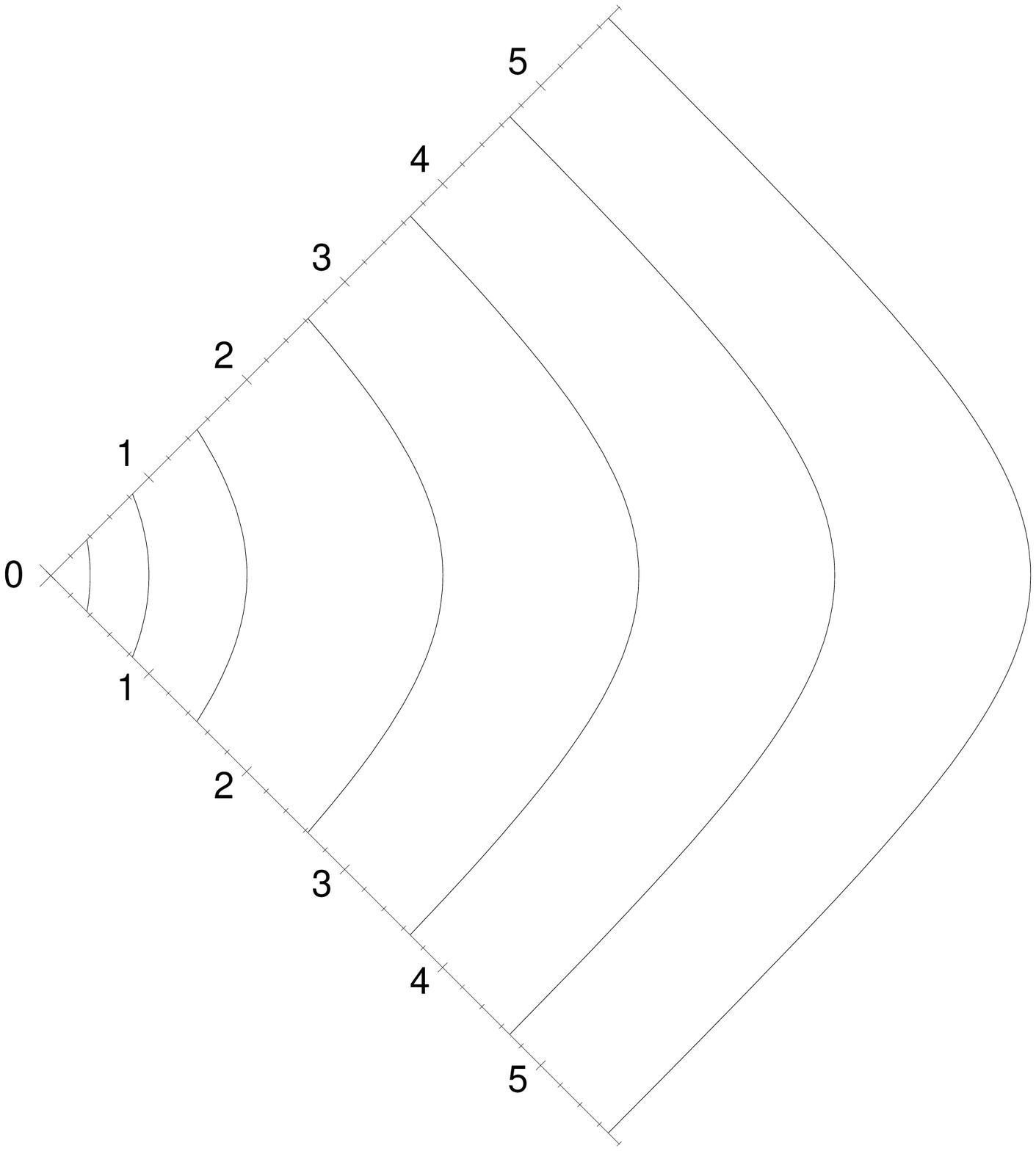}\hfil}}
\vskip\baselineskip
\noindent Figure 4: The flow of $\Gamma_{W}(\tau)$ in a space like wedge.
The whole pattern is invariant under translations in the 
$x^{1}$-direction.
\vfil
}

\vfill\eject
\end